\documentclass[hidelinks,onefignum,onetabnum]{siamart220329}

\usepackage{cite}

\usepackage{cleveref}
\usepackage{lipsum}
\usepackage{amsfonts}
\usepackage{graphicx}
\usepackage{epstopdf}
\usepackage{algorithmic}
\usepackage{tabularray}
\usepackage{makecell}

\newsiamremark{remark}{Remark}
\newsiamremark{hypothesis}{Hypothesis}
\crefname{hypothesis}{Hypothesis}{Hypotheses}
\newsiamthm{claim}{Claim}

\headers{A Weighted-Median Opinion Model on Networks}{L. Mohr, P. G. Hjorth, M. A. Porter}

\title{
    {
    Long-Time and Short-Time Dynamics in a Weighted-Median Opinion Model on Networks}\thanks{
        Submitted to the editors DATE.
    \funding{
        LM acknowledges funding from the Villum Foundation (through the Nation-Scale Social Networks project). PGH acknowledges funding from the Fog Research Institute (contract number FRI-454). MAP acknowledges funding from the National Science Foundation (grant number 1922952) through the Algorithms for Threat Detection (ATD) program.
    }
    }
}

\author{Lasse Mohr\thanks{Department of Applied Mathematics and Computer Science, Technical University of Denmark, Lyngby, Denmark 
  (\email{lmmi@dtu.dk}).}
\and Poul G. Hjorth\thanks{Department of Applied Mathematics and Computer Science, Technical University of Denmark, Lyngby, Denmark 
  (\email{pghj@dtu.dk}).}
\and Mason A. Porter\thanks{Department of Mathematics, University of California, Los Angeles, CA, United States of America; Department of Sociology, University of California, Los Angeles, CA, United States of America; Santa Fe Institute, Santa Fe, NM, United Staes of America (\email{mason@math.ucla.edu})}}

\usepackage{amsopn}

\ifpdf
\hypersetup{
  pdftitle={
  Long-Time and Short-Time Dynamics in a Weighted-Median Opinion Model on Networks
  },
  pdfauthor={L. Mohr, P. G. Hjorth, M. A. Porter}
}
\fi

\begin{document}

\maketitle

\begin{abstract}
Social interactions influence people's opinions.
In some situations, these interactions eventually yield a consensus opinion; in others, they can lead to opinion fragmentation and the formation of different opinion groups in the form of ``echo chambers''.
Consider a social network of individuals with continuous-valued scalar opinions, and suppose that they can change their opinions when they interact with each other.
In many models of the opinion dynamics of individuals in a network, it is common for opinion updates to depend on the mean opinion of interacting individuals.
As an alternative, which may be more realistic in some situations, we study an opinion model with an opinion-update rule that depends on the weighted median of the opinions of interacting individuals.
Through numerical simulations of our median-update opinion model, we investigate how the final opinion distribution depends on network structure.
For configuration-model networks, we also derive a mean-field approximation for the asymptotic dynamics of the opinion distribution when there are infinitely many individuals.
We numerically investigate its accuracy for short-time opinion dynamics on various networks.
\end{abstract}




\section{Introduction}

The opinions of people play important roles in society \cite{granov2017VotersTurnOut, Biddlestone2020ConspiracyCovid}, and the influence that people exert on each other through their social interactions affect these opinions \cite{Lazarsfeld, Bedson2021SocialDiseaseModel}. However, it is difficult to determine the global opinion landscape that arises from such interactions.
In the study of opinion dynamics, researchers investigate how empirically observed phenomena like polarization (the formation of 2 distinct opinion clusters), fragmentation (the formation of 3 or more distinct opinion clusters), and radicalization can emerge from individual-level social mechanisms (such as cognitive dissonance) \cite{Noorazar2020FromDynamics,starnini2025}. 

The concept of cognitive dissonance from social psychology gives some insight into opinion dynamics \cite{Festinger1957}.
According to the notion of cognitive dissonance, individuals experience psychological stress (so-called ``cognitive dissonance") when they disagree with each other and individuals change their opinions to relieve such stress \cite{Festinger1959a}.
Researchers have used this notion to explain a variety of observations about social influence and shifts of political opinions \cite{Beasley2001,Matz2005a}.

When studying large groups of interacting individuals, it is infeasible to directly measure small-scale social {``}forces{"} (such as opinion-shifting effects of cognitive dissonance) that they exert on each other.
Instead, researchers typically investigate how individual-level mechanisms, which one can model using \textit{opinion-update rules}, produce observed large-scale phenomena (such as polarization)~\cite{galesic2021}.

There are numerous models of opinion dynamics~\cite{Xia2011OpinionDynamics, Noorazar2020FromDynamics,starnini2025}.
In opinion models, a typical assumption is that individuals use simple heuristics when navigating complex social domains. 
Individuals in a network that are connected to each other through a social tie can influence each other when they interact.
Traditionally, one models a network as a time-independent graph, which model pairwise (i.e., dyadic) social ties~\cite{newman2018}.
However, opinion dynamics have also been studied on more general network structures, such as temporal networks (which allow social ties to change with time) \cite{KashinMasonTieDecay, EEti2020Centrality-based} and hypergraphs (which incorporate polyadic social ties)~\cite{iacopini2019,HickockOpinDynHypergraphs2022}.


\subsection{Opinion models with continuous-valued opinions}

In the present paper, we suppose that individuals have continuous-valued opinions.
(It is also common to study models, such as voter models~\cite{Redner2019Reality-inspired}, with discrete-valued opinions.) 
For simplicity, we suppose that opinions are real-valued scalars, but one can also let opinions take values in higher-dimensional spaces.
An example situation in which continuous-valued opinions seem appropriate is the choice of the best distribution of state funds to divide between crime prevention and law enforcement, a complex issue that allows opinions that lie somewhere between the two extremes of exclusively funding crime prevention or exclusively funding law enforcement.

There is a wealth of prior research on opinion models with continuous-valued opinions~\cite{Noorazar2020FromDynamics,starnini2025}.
These studies include examinations of whether or not the individuals in a network eventually reach a consensus opinion \cite{Schawe2021WhenSocieties, Lorenz2010HeterogeneousConsensus}, the time to reach a steady state~\cite{Meng2018OpinionNetworks, DingReachingConsensus2019}, public versus private opinions~\cite{Leon-Medina2020Fakers}, phase transitions between regimes of qualitatively different behaviors~\cite{Baron2021ConsensusDisorder, fennell2021, Lorenz2006ConsensusConfidence}, and many other phenomena.
In many opinion models with continuous-valued opinions, the opinion-update rule depends on the mean of the opinions of interacting individuals~\cite{starnini2025}.
In such models, interacting individuals consider a weighted mean of their current opinions, and they then update their opinions if their opinion and the weighted-mean opinion are sufficiently close to each other.
Well-known opinion models with mean-based opinion updates include DeGroot consensus models~\cite{Degroot1974, Jia2015} and bounded-confidence models~\cite{bernardo2024}.


\subsection{Median-based models of opinion dynamics}\label{sec:median_based_models}
Mei et al.~\cite{MeiMicro-Foundation2022, mei2022convergence} recently introduced an opinion model that draws inspiration from cognitive-dissonance theory.
The individuals in their model update their opinions based on median opinions, and they thereby are able to minimize their experienced dissonance.
In their model, Mei et al. supposed that individuals interact with each other in a weighted and directed network. 
The nodes of a network represent the individuals, who have continuous-valued scalar opinions and update their opinions asynchronously.
At each discrete time, a uniformly random individual (i.e., a node) experiences some amount of cognitive dissonance and updates its opinion to a new opinion to minimize it.
This cognitive dissonance is a weighted sum of cognitive-dissonance components from each of that node's neighbors.
The influence weights are proportional to the edge weights, and the cognitive-dissonance contribution from one neighbor is equal to the absolute value of the difference between the node's opinion and that neighbor's opinion.
Using game-theoretic arguments, Mei et al.~\cite{mei2022convergence} showed that minimizing the experienced cognitive dissonance entails individual shifting their opinion to weighted medians of their neighbors' opinions when they update their opinions.

Mei et al. \cite{MeiMicro-Foundation2022} also introduced several variants of their median-based model. These variants include synchronously updating weighted-median models both with self-appraisal (which they called ``inertia") and without self-appraisal.
When self-appraisal is present, individuals consider both their own current opinions and other individuals' current opinions in their opinion updates.
Using data from an earlier online experiment~\cite{Kerckhove2016}, Mei et al. found that a synchronous-update weighted-median model with self-appraisal was more accurate at explaining opinion shifts under social influence than a corresponding weighted-mean model.
Therefore, a weighted-median opinion-update mechanism is sometimes more appropriate than a weighted-mean update mechanism when modeling opinion dynamics.
Aside from their efforts at empirically validating weighted-median mechanisms, Mei et al. \cite{MeiMicro-Foundation2022} did not further study synchronous-update weighted-median models (either with or without self-appraisal).
Therefore, in the present paper, we refer to the asynchronously updating weighted-median opinion model without self-appraisal as ``Mei et al.'s weighted-median opinion model".

Mei et al.'s weighted-median opinion model is a fascinating alternative to opinion models, such as DeGroot models and bounded-confidence models, with update rules that depend on the mean opinions of neighboring nodes~\cite{bernardo2024}.
Subsequent research has included {investigations} of a weighted-median opinion model with continuous time \cite{Han2024continuoustime}, a weighted-median model on unweighted networks \cite{LiWeightedMedian2022}, median-based opinion models with content recommendations \cite{lee2024}, and a weighted-median model with synchronous opinion updates and nodes that are predespositioned to specific opinions~\cite{ZhangWMMPrejudice2025}.

Mei et al.'s weighted-median opinion model has several useful and interesting features, but it also has some potential shortcomings. 
One peculiar feature of their model is that the individuals in a network do not compromise.
Instead, when they are selected to update their opinion, they simply assimilate to their surroundings and replace their opinion with a weighted median of their neighbors' opinions.
Allowing nodes to compromise between their own opinions and their neighbors' opinions can improve the accuracy of weighted-median opinion models at predicting opinion updates~\cite{MeiMicro-Foundation2022}.
For some scenarios, one can also argue that it is undesirable that only a single individual updates their opinion in one time step.
To study group interactions, it may be desirable to use a model that allows multiple individuals to simultaneously update their opinions.

In the present paper, we analyze a synchronously updating analogue of the weighted-median opinion model of Mei et al.~\cite{MeiMicro-Foundation2022, mei2022convergence}.
It is important to study this analogue, as one can obtain very different dynamics with asynchronous and synchronous opinion updates~\cite{Caron-Lormier2008}.
Additionally, using synchronous opinion updates can simplify eventual empirical validations of median-update opinion models, as one can measure human choices (which reflect their underlying opinions) synchronously at discrete times.
It is difficult to empirically validate opinion models~\cite{mas2019}, and accordingly it is important to study opinion models with synchronous node updates.
We also choose to incorporate self-appraisal, as this allows the nodes of a network to not only assimilate the opinions of their neighbors (as in the model of Mei et al.) but also to compromise with their neighbors.
Social psychologists emphasize the importance of self-appraisal to opinion evolution in humans~\cite{TormalaAttituide2018}, and accordingly it has been incorporated in many opinion models, such as DeGroot models~\cite{Degroot1974, Jia2015}, Friedkin--Johnson models \cite{friedkin2014}, and others~\cite{Noorazar2020FromDynamics,starnini2025}.

We investigate the following two questions: 
\begin{enumerate}
	\item{How does the final opinion distribution --- which describes long-time scenarios and approximates the \emph{limit opinion distribution}, which occurs in the infinite-time limit --- depend on network structure and the initial opinion distribution?} 
    \item{Can we effectively describe the short-time dynamics of our model using a mean-field approximation?}
\end{enumerate}


\subsection{Organization of the paper}\label{org}

Our paper proceeds as follows. In \cref{sec:network_int}, we review a few definitions from network science. 
In \cref{sec:WMM}, we define our weighted-median model of opinion dynamics.
In \cref{sec:limit_opinion_dist}, we examine the final opinion distributions of our model for a variety of networks.
In \cref{sec:MF}, derive a mean-field approximation of our weighted-median opinion model and then examine its accuracy. 
Finally, in \cref{sec:disc_and_conc}, we conclude and discuss our findings.
In \cref{sec:limit_opinion_app}, we investigate how the final opinion distribution depends on the parameters of our opinion-update rule and on network structure.
In \cref{sec:mean_field_app}, we {give details about the derivation of}
our mean-field approximation and examine finite-size effects.


\section{A weighted-median model of opinion dynamics}

We consider a weighted-median opinion model with synchronous opinion updating.
We represent each individual as a node of a weighted and directed graph, which encodes the relationships between individuals.
In our opinion model, we suppose that all of these relationships (which are encoded by the edges of a graph) are mutual.


\subsection{Some elementary definitions about networks}\label{sec:network_int}

We now review some relevant definitions and terminology about networks. See \cite{newman2018} for more details.

The simplest type of network is a graph, which is a pair $G = (V, E)$, where $V$ is a set of nodes and $E$ is a set of edges.
In a directed graph, each edge has a direction.
The directed edge $(i,j)$ emanates from a source node $i$ and terminates at a target node $j$.
Because we assume that all edges are mutual, a graph includes the edge $(i,j)$ if and only if it also includes $(j,i)$.
To consider a weighted graph, we assign a positive real value to each edge in $G$.
Such an edge weight, which can be different for the edge from $i$ to $j$ and the edge from $j$ to $i$, can encode features such as a social-connection strength or a communication frequency.
The weight of the edge $(i,j)$ encodes the influence of node $j$ on node $i$.
If nodes $i$ and $j$ are connected by an edge $e$, then the nodes are ``adjacent'' to each other and are neighbors in the graph $G$.
The edge $e$ is ``incident'' to nodes $i$ and $j$.
If $e$ emanates from $i$ to $j$, then $j$ is an ``out-neighbor'' of $i$.
The out-degree (respectively, in-degree) of a node is equal to the number of edges that emanate from (respectively, end at) that node.

An ``undirected walk'' on a directed network is a sequence of nodes in which there is an edge (in any direction) between consecutive nodes in the sequence.
Two nodes $i$ and $j$ are ``weakly connected'' if there is an undirected walk between them.
A ``largest weakly-connected component'' of a directed network is a maximal subset of nodes, along with their associated edges, in which all of the nodes are pairwise weakly connected.
If that component is an entire network, we say that the network is ``weakly connected".

Each weighted, directed network $G$ has an associated ``influence matrix" $W \in [0,1]^{|V|}$ \cite{Degroot1974}.
When there is a directed edge from node $i$ to node $j$, the matrix entry $W_{ij} \in [0, 1]$ gives its weight.
Therefore, $W_{ij}$ encodes the amount of influence of node $j$ on node $i$.
If there is no edge from $i$ to $j$, then $W_{ij} = 0$.
We suppose that $W$ is row-stochastic, so the entries of each of its rows sum to $1$.
This implies that all nodes have nonzero out-degrees, so each node in $G$ is influenced by at least one other node.
When it is convenient, we use $N = |V|$ to denote the number of nodes.


\subsection{Synchronous weighted-median opinion updates}\label{sec:WMM}

Each node $i \in V$ has a time-dependent opinion $x_i(t) \in [0, 1]$ at time $t \in \mathbb{N}_0 = \{0, 1, 2, \ldots \}$.
Recall that $N = |V|$ is the number of nodes. The opinion state $x(t) \in [0,1]^{{N}}$ consists of all opinions $x_i(t)$ for all $i \in V$. Given an initial state $x(0)$, opinions update according to the map
\begin{equation}\label{eq:WMMI}
    x_i(t + 1) = (1 - s)x_i(t) + s\, \mathrm{Med}_i(x(t);W)\,, \quad i \in V\,,
\end{equation}
where $s \in (0,1)$ is the amount of self-appraisal and $\mathrm{Med}_i(x(t);W)$ denotes the weighted median of the opinion states $x(t)$ with respect to the weights $W_{i1}, \, W_{i2},\, W_{i3}, \, \ldots , \,W_{iN}$.
More precisely, 
$\mathrm{Med}_i(x(t);W) = x^{*}\in \mathbb{R}$ if $x^{*}$ satisfies
\begin{equation}\label{Theory:def:weighted_median}
    \sum_{ \{j\, : \, x_j < x^{*} \}} W_{{i}j} \, {\leq}\, \frac{1}{2}  
    \quad \text{and} \quad
    \sum_{\{j\, : \, x_j > x^{*}\}} W_{{i}j}\, {\leq}\, \frac{1}{2}\:\,.
\end{equation}
If multiple values of $x^{*}$ satisfy equation \cref{Theory:def:weighted_median}, we let $\mathrm{Med}_i(x(t);W)$ be the value that is closest to $x_i(t)$.
If two values of $\mathrm{Med}_i(x(t);W)$ are equally close to $x_i(t)$, then we select the smaller of those two values. This notion of a weighted median differs from the more common arithmetic median, which equals the mean of the values that satisfy \cref{Theory:def:weighted_median}. The above weighted median is also valid for studying opinion dynamics, as it yields a unique value for $\mathrm{Med}_i(x(t);W)$ (after applying our tie-breaking procedure, if necessary)~\cite{MeiMicro-Foundation2022}.

We refer to equation \cref{eq:WMMI} as our \textit{opinion-update rule} and refer collectively to equations (\ref{eq:WMMI}, \ref{Theory:def:weighted_median}) as our \textit{weighted-median opinion model}. 
See \cref{Int:fig:WMM_example} for an illustration of how a single node updates its opinion {with the opinion-update rule \cref{eq:WMMI} for a given influence matrix $W$.

Because $W$ is row-stochastic (by assumption), all nodes have nonzero out-degrees.
Therefore, there are no isolated nodes and all nodes have a least one neighbor with an opinion that satisfies the inequalities in \cref{Theory:def:weighted_median}.
Additionally, we consider only weakly connected networks.
If a network is not weakly connected, then one can separately investigate the dynamics of the weighted-median opinion model on each network component.
For simplicity, we assume that all nodes have the same self-appraisal value.

\begin{figure}[t]
    \centering

    \begin{minipage}[t]{.36\textwidth}
      \centering
      (a)\\[12pt]
      \small
      \begin{equation*}
        W=\begin{bmatrix}
          0 & 0.4 & 0.2 & 0.3 & 0.1\\
          0.8 & 0 & 0.2 & 0 & 0\\
          0 & 0.8 & 0 & 0.2 & 0\\
          0 & 0 & 0 & 0 & 1\\
          0.5 & 0 & 0 & 0.5 & 0
        \end{bmatrix}
      \end{equation*}
    \end{minipage}\hspace{9pt}%
    \begin{minipage}[t]{.3\textwidth}
      \centering
      (b)\\[3pt]
      \includegraphics[width=\linewidth]{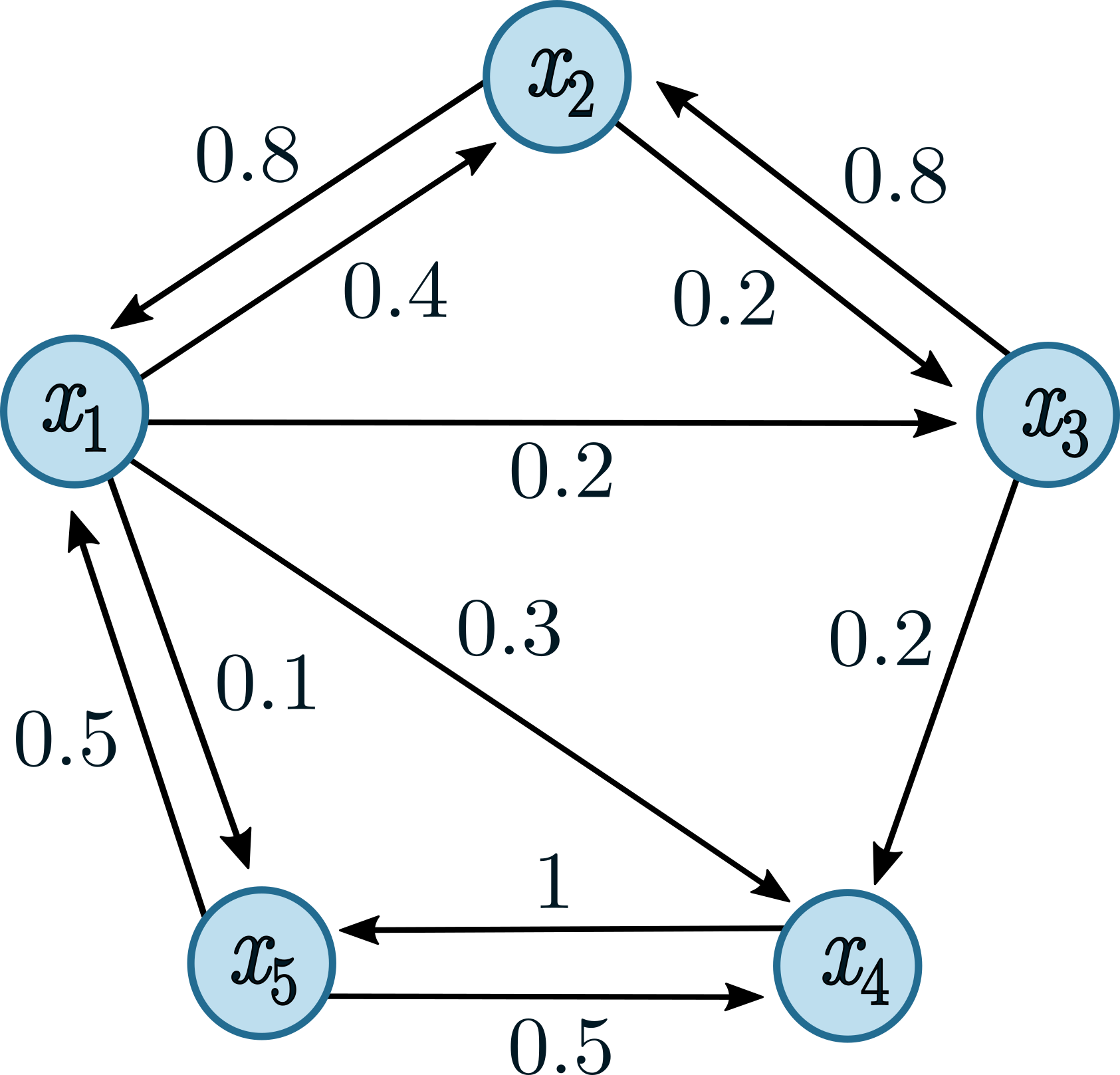}
    \end{minipage}\hspace{9pt}
    \begin{minipage}[t]{.26\textwidth}
      \centering
      (c)\\[12pt]
      \includegraphics[width=\linewidth]{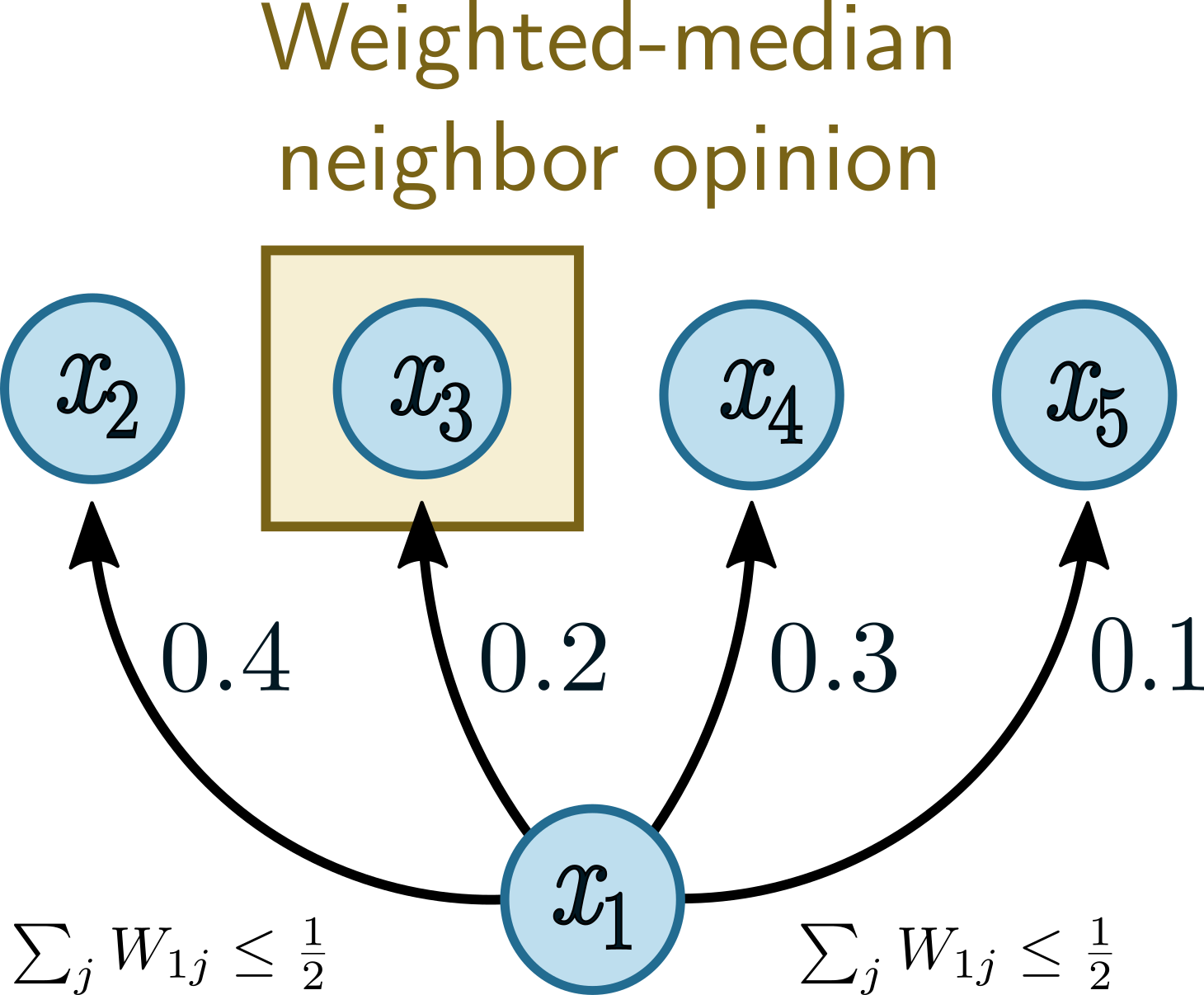}
      \small $x_2 < x_3 < x_4 < x_5$
    \end{minipage}
    \caption{An illustration of the weighted-median opinion-update rule \cref{eq:WMMI}.  
    (a) An example of an influence matrix and (b) its corresponding weighted and directed network.
    (c) Node $1$ updates its opinion by adopting the weighted-median neighbor opinion of its neighborhood.
    Because the update rule \cref{eq:WMMI} is synchronous, all nodes simultaneously update their opinions in this way at each discrete time $t$.
    [Panels (a) and (b) are inspired by Figure 1 of \cite{MeiMicro-Foundation2022}.] 
    }
    \label{Int:fig:WMM_example}
\end{figure}


\section{Final opinion distributions of numerical simulations of our weighted-median opinion model}\label{sec:limit_opinion_dist}

Given a network $G$ and an initial opinion state $x(0)$, it is difficult to analytically determine the opinion distribution as time $t \rightarrow \infty$ (i.e., the limit opinion distribution) of the weighted-median opinion model (\ref{eq:WMMI}, \ref{Theory:def:weighted_median}).
Therefore, to examine the time evolution of opinion distributions and ensuing approximations (so-called ``final opinion distributions") of the limit opinion distributions, we simulate the weighted-median opinion model on a variety of different networks and for a variety of different initial opinion distributions.

The results of our simulations suggests that there are at least two distinct types of limit opinion distributions that can occur in our weighted-median opinion model: (1) a regime in which most opinion values cluster around a single value (and which may correspond to a consensus state) and (2) a regime in which the opinion values spread across opinion space.


\subsection{Simulation specifications}

We do not know if a limit opinion distribution exists for all networks $G$ and all initial opinion states $x(0)$. 
Even when we know that a limit opinion distribution exists, it may take infinitely long to reach it.
Therefore, in practice, we examine approximate limit opinion distributions in our numerical simulations.
We refer to these distributions as ``final opinion distributions".
To obtain the final opinion distributions, we use a convergence criterion.
We interpret the weighted-median opinion model {(\ref{eq:WMMI}, \ref{Theory:def:weighted_median})} as having reached its final opinion distribution at time $t$ if
\begin{equation}\label{sim:conv_critera}
    \max_{i\in V} | x_i(t) - x_i(t + 1)| < \varepsilon\, ,
\end{equation}
with $\varepsilon = 4.44 \times 10^{-16}$ for the examined synthetic networks and $\varepsilon = 10^{-6}$ for the examined real-world networks.
In all of our numerical experiments, the condition \cref{sim:conv_critera} is satisfied in fewer than $10^6$ time steps. 
 
For each network and each initial opinion distribution, we sample the initial opinion of each node randomly from the distribution.
We then update the node opinions at each discrete time using the update rule \cref{eq:WMMI} until we satisfy the convergence criterion \cref{sim:conv_critera}. 
After convergence, we round all opinions to the $14$th decimal place for our simulations on synthetic networks and to the $4$th decimal place for our simulations on real-world networks.
We then count the number of distinct opinions.


\subsection{Network structures}\label{Sec:initial_sim_1}
To develop some understanding of the effects of network structure and the initial opinion distribution on the final opinion distribution, we study our weighted-median opinion model on a variety of networks with a variety of initial opinion distributions. 

We first consider three deterministic synthetic networks: directed versions of cycle networks, prism networks, and square-lattice networks.
In Table \ref{table:synthetic_networks}, we define and give examples of these networks.

\begin{table}[h!]
    \centering
        \begin{tabular}{l | c | c}
        \hline
        Network & Definition & Example \\ 
        \hline \hline \\
        Cycle & 
        \thead{
        For an integer $N \geq 3$, the $N$-node \emph{cycle} network has the node set $\{v_j \, | j\in \{1, \ldots, N\}\}$, 
        \\ directed edges $(v_j, v_{j + 1})$ and $(v_{j + 1}, v_{j})$ for $j\in \{1, \ldots,  N - 1\}$,
        \\and directed edges $(v_N,  v_1)$ and $(v_1, v_N)$.
        }
        &
        \raisebox{-.5\height}{\includegraphics[width=.15\linewidth]{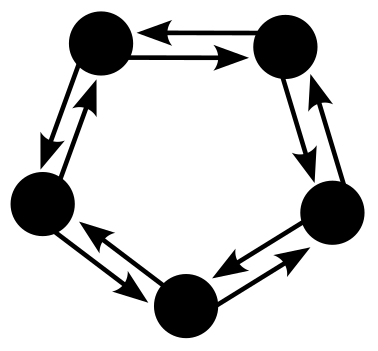}}
        \\ 
        Prism 
        &
        \thead{
        For an even integer $N\geq 6$, let $\{v_j \, | j\in \{1, \dots ,  \frac{N}{2}\}\}$ and $\{u_j \, | j\in \{1, \ldots , \frac{N}{2}\}\}$
        \\be the node sets of two cycles. 
        The $N$-node \textit{prism} network is the union of the two cycles 
        \\with the edges $(v_j, u_j)$ and $(u_j, v_j)$ for $j\in \{1, \ldots , \frac{N}{2}\}$.
        }
        & \raisebox{-.5\height}{\includegraphics[width=0.2\linewidth]{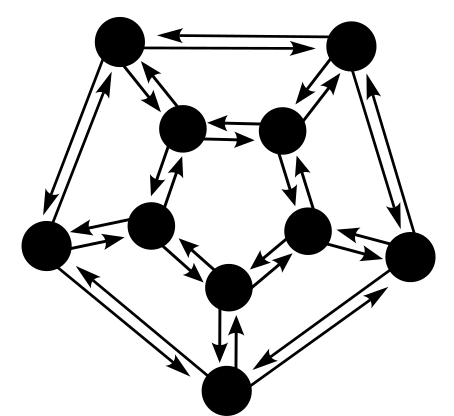}}
        \\
        Square lattice & 
        \thead{
        For integer $k$, the \textit{square lattice} with side length $k$ is the network with node \\set $\{(x,y) \, | \, x,y \in \mathbb{Z}\, \text{ with } \,0 \leq x , y \leq k\}$ and edges $((x_1,y_1),(x_2,y_2))$ and $((x_2,y_2),(x_1,y_1))$ 
        \\for $\|(x_2 - x_1, y_2 - y_1)\|_2 =1$. \\ (The square-lattice network with side length $k$ has $N = (k + 1)^2$ nodes.)
        }
        & \raisebox{-.5\height}{\includegraphics[width=0.15\linewidth]{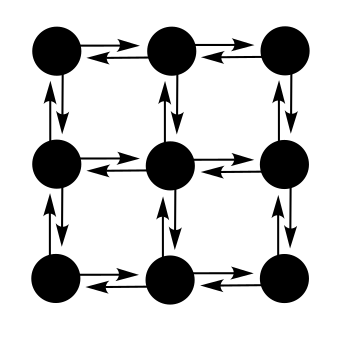}}
        \\ 
        [0.5ex]\\
        \hline
        \end{tabular}
        \caption{The deterministic synthetic networks on which we study our weighted-median opinion model {(\ref{eq:WMMI}, \ref{Theory:def:weighted_median}).}}
    \label{table:synthetic_networks}
\end{table}

We then generate synthetic networks using directed analogues of two well-known random-graph models~\cite{newman2018}: the Barab\'asi--Albert (BA) preferential-attachment model~\cite{Barabasi1999} and the Watts--Strogatz (WS) small-world model~\cite{Watts1998}. 

To construct an $N$-node BA network, we start with a star network with $m \in \mathbb{N}$.
This initial network has $1$ central node and $m - 1$ peripheral nodes, with an undirected edge between the central node and each peripheral node; there are no edges between peripheral nodes.
We grow the network by repeatedly adding a single node and connecting it with an undirected edge to $k$ other nodes, with connection probabilities from linear preferential attachment (i.e., according to the BA mechanism).
We add nodes until there are $N$ nodes in the network.
We then replace all undirected edges by two directed edges, with one in each direction.
We generate a single BA network with $m = 10$, $k = 7$, and $N = 2500${.

Our WS network has $N \in \mathbb{N}$ nodes, which we place in a cycle. 
We start with undirected edges between each node and its $b \in \mathbb{N}$ nearest neighbors.
We then rewire edges in the following manner.
For each edge $(i,j)$, we select one of its indent nodes uniformly at random.
Suppose without loss of generality that we select node $i$.
With probability $p$, we replace the edge $(i,j)$ with an undirected edge $(i,k)$ to a uniformly random node $k \in V \backslash \{i\}$.
We treat any resulting multi-edges as single edges.
We then replace all undirected edges by two directed edges, with one in each direction. 
We generate a single WS network with ${b = 6}$ and $p = 0.1$. 

We also consider real-world networks of Facebook ``friendships'' \cite{Traud2012} and Twitter (which is now called ``$\mathbb{X}$'') ``followerships" \cite{Garimella2018}.
The Facebook networks are undirected, so we replace each edge in them with two directed edges, with one in each direction.

The synthetic networks that we examine each have $N = 2500$ nodes and consist of a single weakly-connected component. 
The opinion-update rule \cref{eq:WMMI} requires each edge to have a weight.
Because we have no prior information about weights for any of the networks, we assume that each source node $i$ is influenced equally by each of its out-neighbors and hence set the weight of each directed edge to $1$ divided by the out-degree of its source node.


\subsection{Initial opinion distributions in our simulations}\label{Sec:initial_sim_2}
To investigate how the final opinion distribution depends on the initial opinion distribution, we consider several different initial distributions. 
Following Mei et al.~\cite{MeiMicro-Foundation2022}, we build these initial opinion distributions using a beta distribution.
Let $\mathrm{Beta}(\alpha, \beta)$ denote the probability distribution with the density function
\begin{equation}
	\mathrm{Beta}(\alpha, \beta) = \frac{\Gamma(\alpha + \beta)}{\Gamma(\alpha)\Gamma(\beta)}x^{\alpha - 1}(1 - x)^{\beta - 1}\,,
\end{equation}
where $\alpha > 0$, $\beta > 0$, and $\Gamma$ is the Gamma function \cite{Pitman1993Probability}.
We consider the following five initial opinion distributions:
\begin{enumerate}
\item{Uniform distribution: we independently sample the initial opinion of each node uniformly at random from the interval $[0,1]$.}
\item{Unimodal distribution: we independently sample the initial opinion of each node from the Beta distribution $\mathrm{Beta}(2,2)$. 
}
\item{Skewed unimodal distribution: we independently sample the initial opinion of each node from the $\mathrm{Beta}(2,7)$ distribution.}
\item{Bimodal distribution: for each node, we sample $X$ from the $\mathrm{Beta}(2,10)$ distribution; we then take the initial opinion of that node to be $X$ with probability $1/2$ and $1 - X$ with probability $1/2$.}
\item{Trimodal distribution: for each node, we sample $X$ from the $\mathrm{Beta}(2,17)$ distribution and sample $Y$ from the $\mathrm{Beta}(12,12)$ distribution; we then take the initial opinion of that node to be $X$, $1 - X$, and $Y$ with probabilities $0.33$, $0.33$, and $0.34$, respectively.}
\end{enumerate}


\subsection{Effect of self-appraisal on the final opinion distribution}
In our numerical simulations, we do not observe any clear dependence of the final opinion distribution on the self-appraisal $s$.
Therefore, we report simulation results only for $s = 0.7$ in the main text. In \cref{App:sec:limit_dist}, we compare the final opinion distributions for several values of self-appraisal on a variety of networks.


\subsection{Final opinion distributions in our simulations}

We now present the results of our simulations of our weighted-median opinion model {(\ref{eq:WMMI}, \ref{Theory:def:weighted_median})} on synthetic networks (see \cref{synthetic}), Facebook friendship networks (see \cref{fb}), and Twitter followership networks (see \cref{twitter}).

We investigate two features of the final opinion distributions: (1) the distribution of opinions and (2) how nodes are organized into ``opinion clusters", which are sets of {pairwise weakly connected} nodes that share the same opinion. For each combination of network and initial opinion distribution, we perform a single simulation until our convergence criterion is satisfied.
We track the number of opinion clusters and the mean, variance, and kurtosis of the opinion-cluster sizes. 
We summarize our observations in \cref{summary}, and we give associated summary statistics in \cref{Res:tab:Lim_dist_group_stat_cycle}--\cref{Res:tab:Lim_dist_group_stat_abortion} in \cref{App:sum_stat_lim_dist}.


\subsubsection{Synthetic networks} \label{synthetic}

We now consider the final opinion distributions that we obtain by simulating our weighted-median opinion model on synthetic networks.
In these simulations, we set the self-appraisal to $s = 0.7$.
We summarize our findings about the size distributions of final opinion clusters in this subsubsection, and we give associated summary statistics in \cref{Res:tab:Lim_dist_group_stat_cycle}--\cref{Res:tab:Lim_dist_group_stat_BA} in \cref{App:sum_stat_lim_dist}.

\begin{figure}[h!]
    \centering
    \includegraphics[width = 0.95 \textwidth]{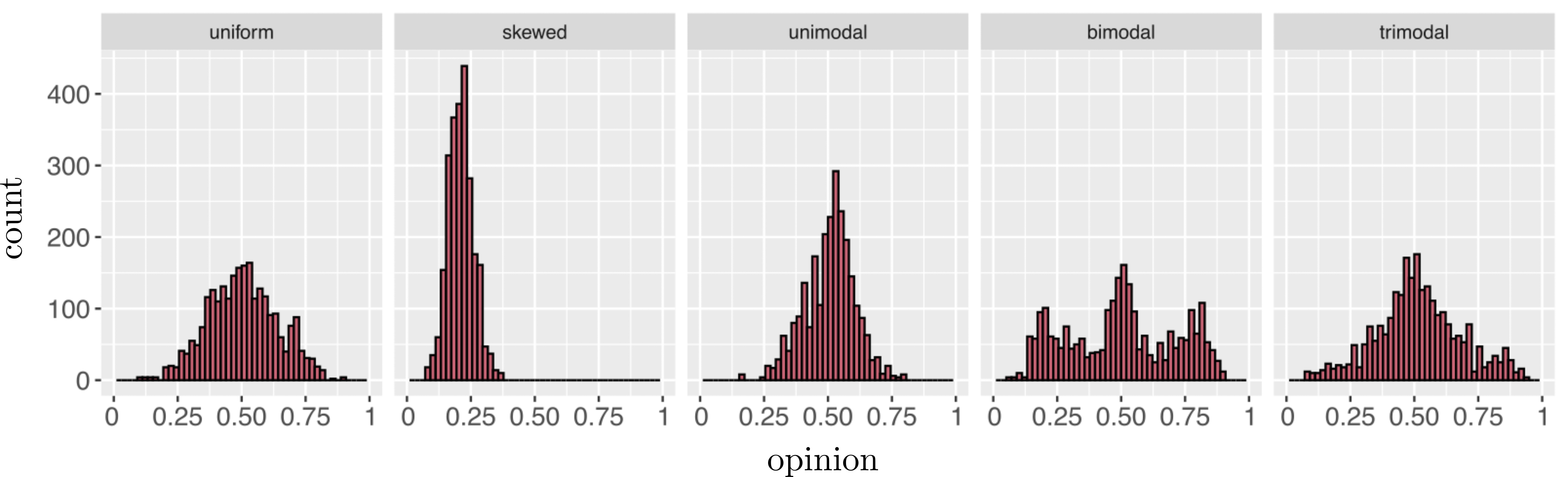}
    \caption{
    Final opinion distributions of our weighted-median opinion model {(\ref{eq:WMMI}, \ref{Theory:def:weighted_median})} with self-appraisal $s = 0.7$ for various initial opinion distributions on a cycle network with $2500$ nodes.
    Each histogram corresponds to a single final opinion distribution that starts from the indicated initial opinion distribution. 
    We obtain similar final opinion distributions for other values of self-appraisal.
    }\label{res:fig:cycle_WMMI_limit_dist}
\end{figure}

We first examine the final opinion distributions of our weighted-median opinion model with the uniform, unimodal, skewed, bimodal, and trimodal initial opinion distributions (see \cref{Sec:initial_sim_2}) for several synthetic networks.
For each combination of network and initial opinion distribution, we perform a single simulation until our convergence criterion is satisfied.
We then examine the resulting final opinion distribution.
To illustrate these distributions, in \cref{res:fig:cycle_WMMI_limit_dist}, we show histograms of the final opinion distribution for simulations on the cycle network.

For the cycle, the prism, and the square-lattice network, we obtain similar final mean opinion-cluster sizes for the different initial opinion distributions.
They are much smaller than the network size and range from $2.48$ to $3.21$ across our simulations on the three networks. 
However, the kurtoses of the opinion-cluster sizes for the square-lattice network, which range from $26.59$ to $55.15$ in our simulations, are notably larger than for the cycle and prism networks, for which the observed kurtoses range from $0.9$ to $3.29$.

For our WS network, we observe that the number of final opinion clusters, the means of the distributions of the opinion-cluster sizes, and the variances of the cluster-size distributions are similar for all examined initial opinion distributions.
The opinion-cluster size kurtoses, which range from $5.70 \times 10^3$ to $1.52 \times 10^4$, for our WS network are much larger than the kurtoses for the prism, grid, and cycle networks.
The large kurtoses imply that the sets of opinion-cluster sizes have many outliers, extreme outliers, or both. The opinions of these outliers must be larger than the mean opinion because the mean opinion-cluster size, which ranges from $6.02$ to $6.67$, is much smaller than the network size. Therefore, the large kurtoses indicates that the final opinion distributions for our WS network have larger opinion clusters than those in the prism, cycle, and square-lattice networks. 

For our BA network, the final cluster-size kurtoses, which range from $2.17 \times 10^{11}$ to $3.54 \times 10^{11}$, are even larger than those for our WS network.
Additionally, the variances of the opinion-cluster sizes are much larger for our BA network than for our WS network.
For our WS network, the largest observed variance of the opinion-cluster size is $27.12$, but our BA network has variances that range from $6.62 \times 10^4$ to $8.9 \times 10^4$.
The large kurtoses indicates that the final opinion distribution for our BA network has opinion clusters that are much larger than the mean opinion-cluster size.

In our simulations, the opinions in the final opinion distribution include opinion values throughout the opinion space for the prism, square-lattice, cycle, and WS networks.
However, for our BA network, for each initial opinion distribution, a small interval of the opinion space contains all of the opinions in the final opinion distribution. 
The length of this interval is about $0.015$, so the final opinion has support on less than $2$\% of the opinion space.
We conclude from our simulations that the final opinion distribution of our weighted-median opinion model is fundamentally different for our BA network than for the other synthetic networks.


\subsubsection{Facebook friendship networks}\label{fb}

We now examine the final opinion distributions of our weighted-median opinion model with the uniform, skewed, unimodal, bimodal, and trimodal initial opinion distributions (see \cref{Sec:initial_sim_2}) on three Facebook friendship networks~\cite{Traud2012}.
In these networks, each node is an individual in one United States university and each edge is a Facebook ``friendship'' between them.
Because we define our weighted-median opinion model on directed networks, we replace each undirected edge with two directed edges, with one in each direction.
The weight of each directed edge is 1 divided by the out-degree of the source node.
We use networks from Caltech (which has $762$ nodes), Bowdoin (which has $2250$ nodes), and Georgetown (which has $9388$ nodes).
For each combination of network and initial opinion distribution, we perform one simulation of our weighted-median {opinion} model {(\ref{eq:WMMI}, \ref{Theory:def:weighted_median})} until our convergence criterion is satisfied.
We then examine the resulting final opinion distribution.
We summarize our findings about the size distributions of final opinion clusters in this subsection, and we give associated summary statistics in \cref{Res:tab:Lim_dist_group_stat_Caltech}--\cref{Res:tab:Lim_dist_group_stat_Georgetown} in \cref{App:sum_stat_lim_dist}.

We observe similar distributions of final opinion-cluster sizes for the three Facebook friendship networks.
The mean opinion-cluster sizes for each of these networks range from $2.55$ to $13.14$, which are much smaller than the network sizes.
The kurtoses of the opinion-cluster sizes are notably large.
For instance, the kurtosis for the Georgetown network ranges from $3.83 \times 10^{8}$ to $2.20 \times 10^9$.
These large kurtoses suggest that the opinion-cluster-size distribution is heavy-tailed.
The variances of the opinion-cluster sizes for the Facebook friendship networks are smaller than those in our BA network.
Therefore, the Facebook networks have a smaller spread than our BA network in the opinion-cluster sizes around the mean.

For each of the Facebook friendship networks and for all initial opinion distributions, a small number of large final opinion clusters include most of the nodes.
Additionally, in the final opinion distributions, a small region of opinion space encompasses all of the opinions in these clusters.
We made the same qualitative observation for the BA network, which also has large final opinion clusters with opinions in a small region of opinion space.

We also observe that approximately $10$\% of the nodes of the Facebook ``Friendship'' networks are in small opinion clusters of the final opinion distribution.
The opinion values of these small opinion clusters are distributed roughly uniformly in opinion space.
We do not observe this combination of a few large opinion clusters with similar opinion values and many small opinion clusters with dissimilar opinion values in the final opinion distributions for any of the synthetic networks.


\subsubsection{Twitter followership networks}\label{twitter}

We now examine the final opinion distributions of our weighted-median opinion model on two Twitter followership networks \cite{Garimella2018}.
In these two networks, each node is a Twitter account that tweeted about a given topic and a directed edge represents a follower relationship between two accounts.
The source account of an edge follows {the edge's} target account.
These two networks involve discussions of the topics Obamacare\footnote{The Affordable Care Act, which is known colloquially as Obamacare, is a controversial United States law that was enacted in 2010 during the presidential administration of Barack Obama. The law was the subject of intense online debate (including on Twitter).} and abortion.
The Obamacare network has $8006$ nodes, and the abortion network has $6114$ nodes.
The weight of an edge is 1 divided by the out-degree of its source node.
For each combination of Twitter followership network and initial opinion distribution, we perform one simulation of our weighted-median opinion model (\ref{eq:WMMI}, \ref{Theory:def:weighted_median}) until our convergence criterion is satisfied.
We then examine the resulting final opinion distribution.
We summarize our findings about the size distributions of final opinion clusters in this subsection, and we give associated summary statistics in \cref{Res:tab:Lim_dist_group_stat_obamacare} and \cref{Res:tab:Lim_dist_group_stat_abortion} in \cref{App:sum_stat_lim_dist}.

We obtain very different final opinion distributions for the Twitter followership networks than for our synthetic networks.
In our simulations on both Twitter followership networks, we observe similar size distributions of the final opinion clusters for all initial opinion distributions. 
The mean final opinion-cluster sizes are much smaller than the network sizes, and the kurtoses are large.
Therefore, as we observed for the final opinion distribution for the BA network and Facebook friendship networks, the final opinion distribution for the Twitter followership networks have heavy-tailed cluster-size distributions and the opinions in the large opinion clusters are in a small region of opinion space.
We also observe that a small fraction of the nodes are in small opinion clusters, which have opinion values that are distributed roughly uniformly in opinion space.
We also observe this combination of a few large opinion clusters and many smaller opinion clusters with dissimilar opinion values for the Facebook friendship networks, but we do not observe it for any of the synthetic networks.


\subsubsection{Summary of our observations}\label{summary}

We group the final opinion distributions of the examine networks into two rough groups: (1) the BA network, the Facebook friendship networks, and the Twitter followerwship networks; and (2) the cycle network, the prism network, the square-lattice network, and the WS network. The networks in the first group have a small number of large final opinion clusters with similar opinion values corresponding to a set of dominant (or hegemonious) opinions, whereas the networks in the second group do not have large final opinion clusters. Taken together, our numerical results suggest that our weighted-median opinion model can yield at least two different qualitative types of limit opinion distributions.
One regime has a dominant opinion, and the other regime does not. 


\section{Mean-field approximation of our model and its short-time dynamics}\label{sec:MF}

We now derive a mean-field approximation of our weighted-median opinion model and study our model's short-time dynamics.


\subsection{Our mean-field approximation}

We suppose that a set of time-dependent probability distributions govern the probability densities of the node opinions at each time step, and we derive difference equations for the time evolution of these densities.
Our derivation is inspired by the mean-field approximation of the Deffuant--Weisbuch model in \cite{fennell2021}.

We are deriving a degree-based mean-field approximation~\cite{PorterGleeson2016}, so we assume that the opinions of all degree-$k$ nodes are statistically identical.
Therefore, the opinions of all degree-$k$ nodes follow a degree-specific opinion distribution $P_k(x,t)$.
We assume that these opinions evolve on a network that is generated by a configuration model with a prescribed degree distribution~\cite{Fosdick2016ConfiguringSequences}.
We also assume that the network is ``annealed''; therefore, at each time step, we rewire the edges of the configuration-model network while preserving the in-degree and out-degree of each node~\cite{Dorogovtsev2008CriticalNetworks}.
In this section, we give an intuitive explanation of the equations in the mean-field approximation.
In \cref{App:MF_derivation}, we give the details of our derivation of this approximation.

Consider the degree distribution $\{q_l\}$.
For each $l \in \mathbb{N}$, let $q_l$ be the probability that a uniform randomly selected node has degree $l$.
Let $v \in V$ be a uniformly random degree-$k$ node at time $t \in \mathbb{N}_0 = \{0, 1, \ldots \}$.
By assumption, node $v$'s opinion follows the distribution $P_k(x,t)$.
To update it, we calculate the weighted-median neighbor opinion of its $k$ neighbors using \cref{Theory:def:weighted_median}.
Because we are considering annealed networks, the opinion distribution of each of these neighboring nodes follows the same distribution asymptotically as the network size (i.e., the number of nodes) tends to infinity.
Let $\phi(x,t)$ denote the opinion distribution of a neighbor of node $v$ at time $t$, and let $\Phi(x,t)$ be the cumulative density function of $\phi(x,t)$.
Employing the annealing assumption, we write
\begin{equation*}
    \phi(x,t) = \sum_l \pi_l P_l(x,t)\,,
\end{equation*}
where $\pi_l$ is the probability that a uniformly random neighbor of $v$ has degree $l \in \mathbb{N}$ as the network size tends to infinity.
The opinion distribution $\phi(x,t)$ is independent of the degree $k$.

In the limit of infinitely many nodes, the opinions of node $v$'s neighbors are $k$ independent samples from the distribution $\phi(x,t)$.
To find the distribution of the weighted median of these $k$ opinions, we use the theory of ``order statistics"~\cite{Sarndal2004OrderStatistics} and thereby obtain that the weighted-median neighbor opinion follows the distribution 
\begin{equation*}
    \theta_k(x,t) = 
		    \begin{cases}
                \frac{{(2m + 1)}!}{m!m!} \phi(x,t) \Phi(x,t)^m \left(1 - \Phi(x,t)\right)^m\,,\, \: \text{$k = 2m + 1$ (i.e., odd $k$)} \\
                \frac{{(2m)}!}{(m - 1)!m!}\phi(x,t)\big(\Phi(x,t) \left(1 - \Phi(x,t)\right)\big)^{m-1} c(x,t)\,,\, \:\text{$k = 2m$ (i.e., even $k$)} \,,
    		\end{cases}
\end{equation*} 
where $c(x,t)$ is a probability distribution that accounts for the possibility there there are two candidates for the weighted-median neighbor opinion when $v$ has an even number of neighbors.
See \cref{App:MF_derivation} for further details about $c(x,t)$.

To update its opinion, node $v$ adjusts its opinion towards the weighted median of the opinions of its neighbors.
The self-appraisal $s$ controls the magnitude of this opinion shift.
Because we only know the distributions of opinions (rather than, e.g., the precise opinion of each node), this opinion shift corresponds to updating the opinion distribution $P_k(x,t)$ of degree-$k$ nodes using the equation
\begin{equation}\label{Theory:eq:mean_field_eq}
    P_k(x,t + 1) = \frac{1}{s}\int_{y\in [0,1]}\theta_k\left( \frac{x + (s - 1) y}{s}, t\right)P_k(y,t) \,\mathrm{d} y\,.
\end{equation}

The above procedure holds for any ``degree class" $k$ (i.e., for all nodes with degree $k$ for each value of $k$).
From the opinion distribution of each degree class, we obtain the global opinion distribution 
\begin{align}\label{Theory:eq:global_dens}
    P(x,t) = \sum_k q_k P_k(x,t)\, .
\end{align}
The {system of equations (\ref{Theory:eq:mean_field_eq}, \ref{Theory:eq:global_dens})} is a mean-field approximation of the weighted-median opinion model (\ref{eq:WMMI}, \ref{Theory:def:weighted_median}).


\subsection{Specifications for testing the accuracy of our mean-field approximation {(\ref{Theory:eq:mean_field_eq}, \ref{Theory:eq:global_dens})}}\label{sec:num_test_MF}

We do not know how to analytically solve {(\ref{Theory:eq:mean_field_eq}, \ref{Theory:eq:global_dens})}.
For a given network and a given initial opinion distribution, we want to investigate how well the mean-field approximation {(\ref{Theory:eq:mean_field_eq}, \ref{Theory:eq:global_dens})} describes the evolution of our weighted-median opinion model (\ref{eq:WMMI}, \ref{Theory:def:weighted_median}).
Therefore, we solve (\ref{Theory:eq:mean_field_eq}, \ref{Theory:eq:global_dens})
numerically and compare our solutions to empirically estimated opinion densities from direct numerical simulations of (\ref{eq:WMMI}, \ref{Theory:def:weighted_median}).

To solve the mean-field approximation numerically, we discretize the opinion space into $1025$ equally spaced values and perform first-order interpolation to evaluate $\theta_k$ between these values.
We obtain numerical solutions of the same accuracy when we instead discretize the space into $2050$ values.
For each network, we calculate the degree distribution $q_k$ using the network's empirical degree distribution.
To ensure numerical stability of our solution, we lump all degree classes with degrees $1000$ or more into a single degree class. 
For further details about this lumping scheme, see \cref{sec:lumping}.

We investigate the time evolution of the accuracy of the mean-field approximation {(\ref{Theory:eq:mean_field_eq}, \ref{Theory:eq:global_dens})} for all of our networks. 
We restrict our analysis to short time scales.
Our mean-field approximation is accurate for short-time dynamics, but we observe empirically that its accuracy deteriorates exponentially with time. 

The mean-field approximation does not account for the direction of edges, so we disregard edge directions when we compute the degree distribution $\{q_k\}$.
We consider three deterministic synthetic networks (complete networks, cycle networks, and prism networks), the BA and WS random-graph models (see \cref{Sec:initial_sim_1} for our choices of parameter values), and networks that we generate using a configuration model \cite{Fosdick2016ConfiguringSequences,newman2018} with nodes of two degree classes, $k_1 = 11$ and $k_2 = 101$, with associated probabilities $q_{k_1} = 0.9$ and $q_{k_2} = 0.1$.
These configuration-model networks are examples of $(k_1,k_2)$-regular random graphs~\cite{melnik2013}.
For each of the three random-graph models, we construct $100$ networks with $2500$ nodes each. We also study three Facebook friendship networks and two Twitter followership networks (see \cref{Sec:initial_sim_1}). 

The accuracy of our mean-field approximation {(\ref{Theory:eq:mean_field_eq}, \ref{Theory:eq:global_dens})} deteriorates with time.
Two important sources of errors stem from our assumptions in our derivations of the approximation.
First, we derive the approximation in the $N \rightarrow \infty$ asymptotic regime.
Second, we assume that the dynamics occur on an annealed network \cite{Dorogovtsev2008CriticalNetworks}. 
In \cref{App:finite-size}, we examine how these assumptions affect the accuracy of our mean-field approximation.


\subsection{Accuracy of the mean-field approximation~{(\ref{Theory:eq:mean_field_eq}, \ref{Theory:eq:global_dens})}}\label{sec:MF_acc}

We explore the accuracy of the mean-field approximation {(\ref{Theory:eq:mean_field_eq}, \ref{Theory:eq:global_dens})} by comparing numerical solutions of it with empirically estimated opinion densities from direct numerical simulations of the short-time dynamics of our weighted-median opinion model.
As we will discuss shortly, the accuracy of the mean-field approximation deteriorates with time in a nontrivial way that depends both on the self-appraisal value and on the network structure.
Accordingly, our mean-field approximation is most 
useful for short-time dynamics.

We perform a variety of numerical experiments.
For our simulations of our weighted-median {opinion} model (\ref{eq:WMMI}, \ref{Theory:def:weighted_median}) on all networks, we sample the initial opinions of the nodes uniformly at random.
For the synthetic networks and the Bowdoin Facebook friendship network, we estimate the opinion distributions by averaging the opinion distributions from $100$ independent simulations with different sets of initial node opinions.
Each simulation that uses a random-network model employs a different network that we construct using that model.
Because of the large network sizes of the Georgetown Facebook network and both Twitter followership networks, we estimate the opinion distributions using $25$ independent simulations with different sets of initial node opinions.
For the Caltech Facebook network, we use $300$ independent simulations with different sets of initial node opinions.

In \cref{Res:fig:MF_accuracy_config}, we plot the natural logarithm of the root-mean-square error (RMSE) between our mean-field approximation {(\ref{Theory:eq:mean_field_eq}, \ref{Theory:eq:global_dens})} and our weighted-median opinion model (\ref{eq:WMMI}, \ref{Theory:def:weighted_median}).
We observe that the RMSE increases with time and that the RMSE increases faster for larger values of self-appraisal.

\begin{figure}[h!]
    \centering
    \includegraphics[width = 0.95 \textwidth]{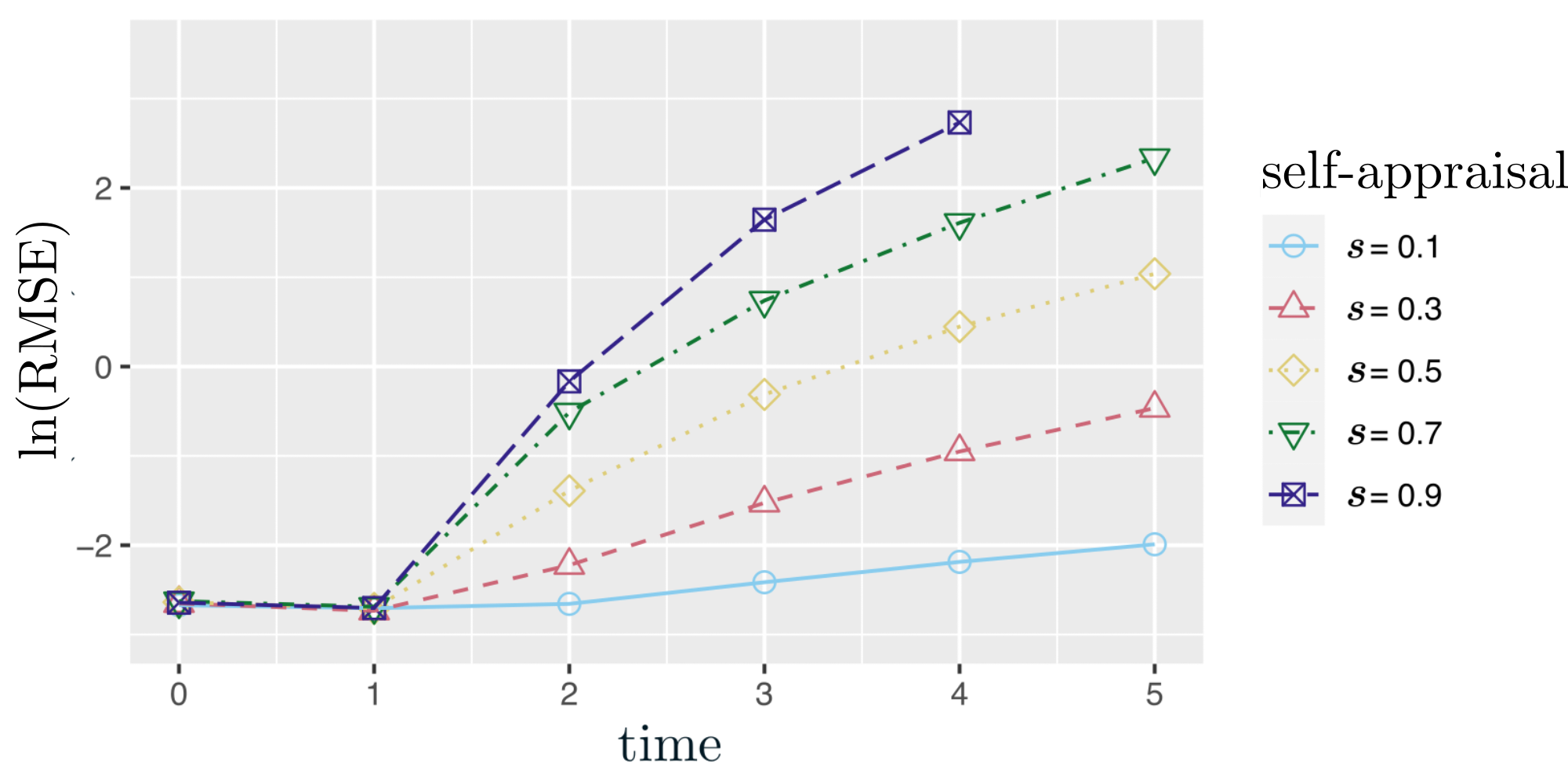}
    \caption{
    The natural logarithm of the root-mean-square error (RMSE) between our mean-field approximation (\ref{Theory:eq:mean_field_eq}, \ref{Theory:eq:global_dens}) and our weighted-median opinion model {(\ref{eq:WMMI}, \ref{Theory:def:weighted_median})} for a variety of self-appraisal values $s$ as a function of discrete time $t$ for $(k_1,k_2)$-regular configuration-model networks with nodes of degrees $k_1 = 11$ and $k_2 = 101$ with associated probabilities $q_{k_1} = 0.9$ and $q_{k_2} = 0.1$.
    At first, the mean-field approximation accurately describes the evolution of the opinion distribution, but the accuracy subsequently deteriorates, with substantial deterioration when $s$ is not small.
    }\label{Res:fig:MF_accuracy_config}
\end{figure}

To investigate why the RMSE error increases faster for larger values of self-appraisal, we examine the sample mean $\overline{P}(x,t)$ of the opinion densities from our weighted-median opinion model (\ref{eq:WMMI}, \ref{Theory:def:weighted_median}) and the numerical solution $\hat{P}(x,t)$ of our mean-field approximation {(\ref{Theory:eq:mean_field_eq}, \ref{Theory:eq:global_dens})} for self-appraisal values of $s = 0.1$ and $s = 0.9$ (see \cref{Res:fig:MF_config}). 
For $s = 0.9$, the empirical distribution and mean-field approximation eventually diverge from each other.
The mean-field approximation evolves towards a Dirac delta function, whereas the empirical opinion distribution retains some spread in the opinion values.
For $s = 0.1$, the mean-field approximation eventually underestimates the opinion spread, but this occurs much less severely than for $s = 0.9$.

\begin{figure}[h!]
    \centering
    \includegraphics[width  = 0.95 \textwidth]{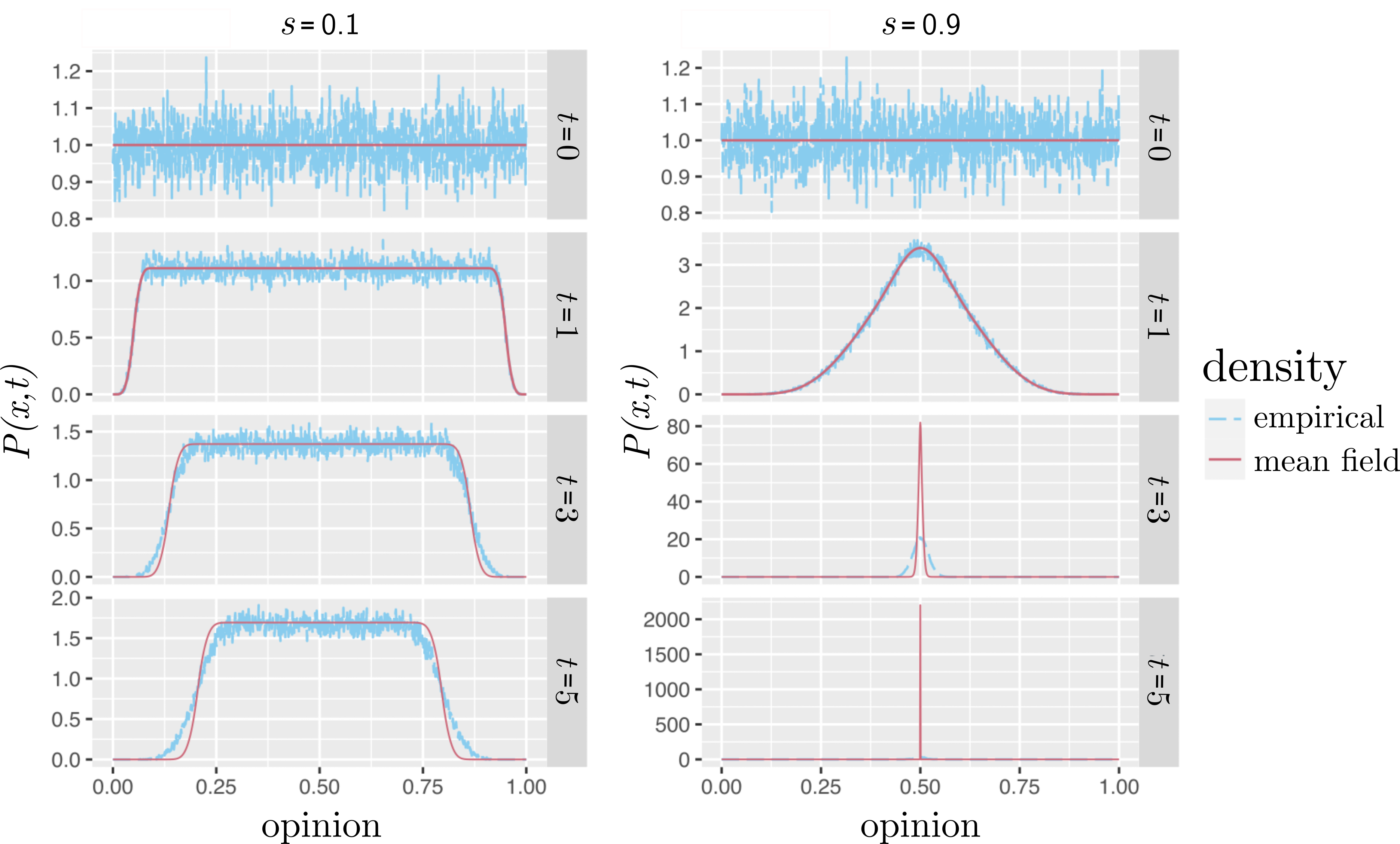}
    \caption{
        Time evolution of the empirical opinion distribution of our weighted-median opinion model {(\ref{eq:WMMI}, \ref{Theory:def:weighted_median})} and the mean-field approximation {(\ref{Theory:eq:mean_field_eq}, \ref{Theory:eq:global_dens})} for self-appraisal values of (left) $s = 0.1$ and (right) $s = 0.9$ on configuration-model networks from (top) time $t = 0$ to (bottom) time $t = 5$. The $(k_1,k_2)$-regular configuration-model networks have nodes with degrees $k_1 = 11$ and $k_2 = 101$ with associated probabilities $q_{k_1} = 0.9$ and $q_{k_2} = 0.1$. 
    For $s = 0.1$, the mean-field approximation gives a reasonable but imperfect approximation of direct numerical simulations of our weighted-median opinion model. 
    However, for $s = 0.9$, the mean-field approximation eventually evolves towards a Dirac delta function, whereas the empirical opinion distribution from our weighted-median opinion model does not.
    }\label{Res:fig:MF_config}
\end{figure}

In \cref{Res:fig:MF_accuracy_cycle}, we plot the natural logarithm of the RMSE between our mean-field approximation {(\ref{Theory:eq:mean_field_eq}, \ref{Theory:eq:global_dens})} and our weighted-median opinion model (\ref{eq:WMMI}, \ref{Theory:def:weighted_median}) for a variety of self-appraisal values $s$ for the cycle network.
In contrast to the RMSE for the configuration-model networks, the RMSE for the cycle network does not increase faster for larger values of self-appraisal. 
Instead, of the examined values of $s$, the RMSE is the largest for $s = 0.5$ and smallest for $s = 0.1$ and $s = 0.9$.
After time $t = 3$, the RMSE increases at a similar rate for all examined values of $s$.
The RMSE behavior for the WS networks is similar to that for the cycle network.
These are the only two of the examined network types for which the maximum observed RMSE occurs at self-appraisal values other than $s = 0.9$.

\begin{figure}[h!]
    \centering
    \includegraphics[width = 0.95 \textwidth]{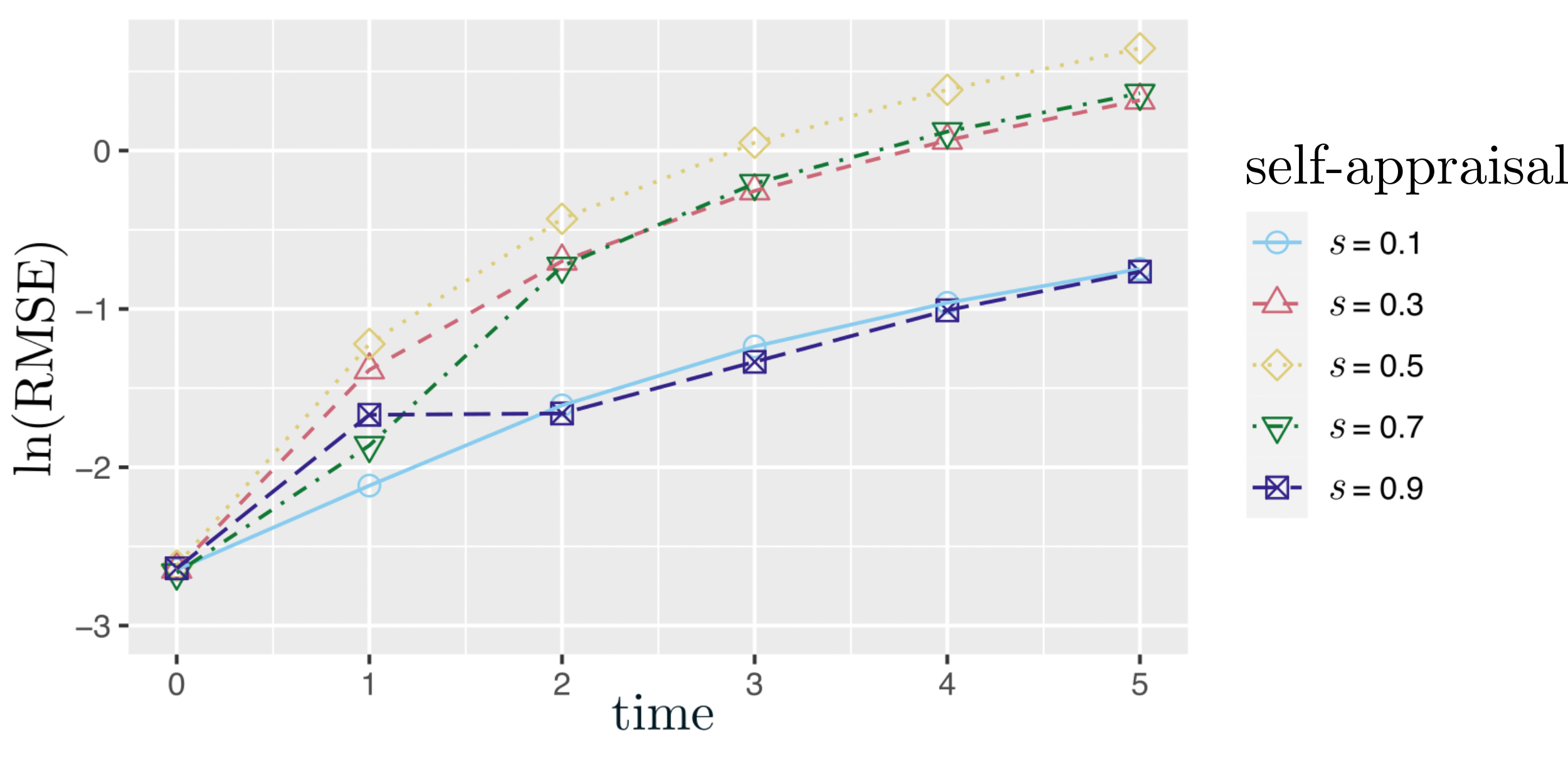}
    \caption{
    The natural logarithm of the root-mean-square error (RMSE) between the mean-field approximation {(\ref{Theory:eq:mean_field_eq}, \ref{Theory:eq:global_dens})} and our weighted-median opinion model {(\ref{eq:WMMI}, \ref{Theory:def:weighted_median})} for cycle networks.
    The accuracy of the mean-field approximation does not depend monotonically on the self-appraisal $s$.
     }
    \label{Res:fig:MF_accuracy_cycle}
\end{figure}

In \cref{Res:fig:MF_accuracy_Obamacare}, we plot the natural logarithm of the RMSE between our mean-field approximation {(\ref{Theory:eq:mean_field_eq}, \ref{Theory:eq:global_dens})} and our weighted-median opinion model (\ref{eq:WMMI}, \ref{Theory:def:weighted_median}) for a variety of self-appraisal values $s$ for the Obamacare Twitter followership network.
For $s \geq 0.3$, the RMSE stays almost the same after one time step before increasing noticeably, with faster increases for larger values of $s$.
For $s = 0.1$, the RMSE increases substantially in the first time step and subsequently increases slowly.
Based on our simulations, we conclude that there is a transition in $s$ between these two qualitatively different behaviors.
We observe similar patterns for the RMSE for all three Facebook friendship networks and both Twitter followership networks.
To examine whether the transition is abrupt or smooth in nature, in \cref{App:sec:MF_RMSE_refined}, we investigate the transitional behavior of the RMSE between $s = 0.1$ and $s = 0.3$ for the Georgetown Facebook friendship network.
Based on these computations, the transition appears to be smooth.

\begin{figure}[h!]
    \centering
    \includegraphics[width  = 0.95 \textwidth]{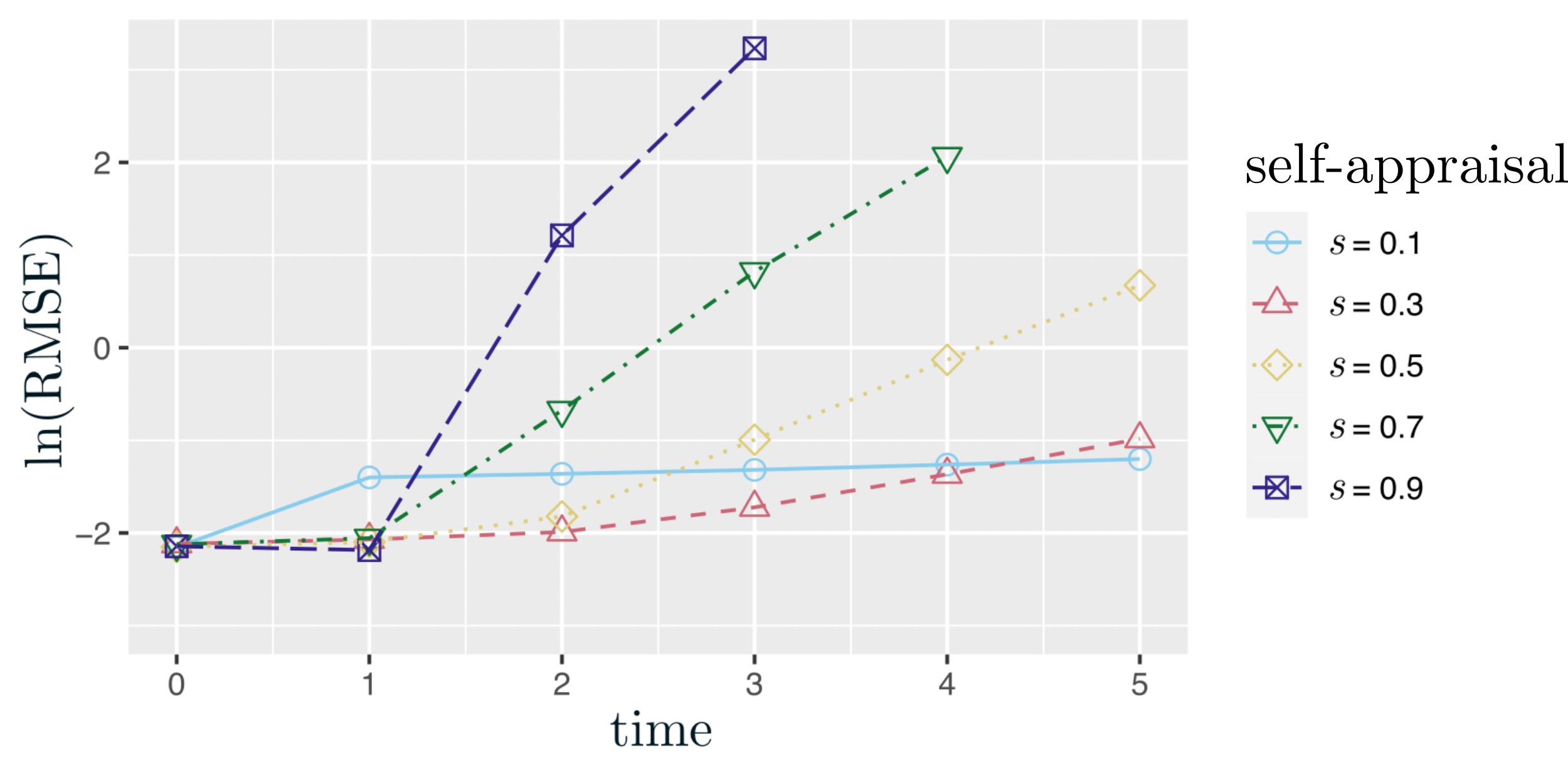}
    \caption{
        The natural logarithm of the RMSE between our mean-field approximation {(\ref{Theory:eq:mean_field_eq}, \ref{Theory:eq:global_dens})} and our weighted-median opinion model for the Obamacare Twitter followership network.
    For small self-appraisal values $s$, the RMSE initially increases notably and subsequently increases slightly.
    By contrast, for larger self-appraisal values, the RMSE is initially almost constant and subsequently increases sharply. We observe a similar pattern for our other real-world social networks.
    }
    \label{Res:fig:MF_accuracy_Obamacare}
\end{figure}

{
From our results, it is difficult to draw major conclusions about the connection between network structure and the accuracy of our mean-field approximation (\ref{Theory:eq:mean_field_eq}, \ref{Theory:eq:global_dens}).
In our numerical experiments, we observe that our mean-field approximation is more accurate for the cycle networks, square lattices, prisms than for BA networks, the Facebook friendship networks, and the Twitter followership networks.
It is possible that the larger-degree nodes in the latter networks amplify the deterioration with time of the accuracy of our mean-field approximation.


\section{Conclusions and discussion}\label{sec:disc_and_conc}

We formulated and studied a synchronously updating analogue of the asynchronously updating weighted-median opinion model of Mei et al.~\cite{MeiMicro-Foundation2022}.
 To understand the long-time dynamics of our weighted-median opinion model, we numerically simulated it on a variety of networks and explored the resulting final opinion distributions and final opinion-cluster-size distributions.
We also derived a mean-field approximation of the evolution of opinion densities, and we examined the short-time accuracy of this approximation.

In our numerical computations, we demonstrated that the amount of self-appraisal has little effect on our model's final opinion distribution (although it does affect trajectories through opinion space), which is influenced much more significantly by network structure.

For example, in the examined Facebook friendship networks, Twitter followership networks, and a Barab\'asi--Albert network, we obtained large opinion clusters and observed that the opinion values of these large clusters are close to each other in opinion space, whereas the opinion values of small opinion clusters are spread throughout opinion space.
By contrast, we did not observe large opinion clusters in the final opinion distribution for the cycle network, the prism network, the square-lattice network, and a Watts--Strogatz network.
Instead, in these final opinion distributions, there are many small opinion clusters with opinion values that are scattered throughout opinion space.
Overall, our numerical observations suggest that our weighted-median opinion model may have at least two regimes in its final opinion distributions.
One of these regimes has a single dominant opinion (and may correspond to a consensus state), and the other does not.
Because we only performed a single simulation for each combination of network and initial opinion distribution, we do not have sufficient statistical power to hypothesize which conditions yield these regimes.
It is worthwhile to conduct further simulation-based and mathematically rigorous investigations to investigate the conditions that yield such qualitatively different behaviors in limit opinion distributions for different types of networks.

In our paper, we also derived and numerically investigated a mean-field approximation of our weighted-median {opinion} model.
We demonstrated with numerical computations that our mean-field approximation accurately describes the dynamics of our model's opinion distributions on short time scales for the examined networks.
However, as we observed in our numerical computations, the accuracy of this approximation deteriorates with time at a rate that depends both on the self-appraisal value and on network structure.
For most of the networks that we considered, our mean-field approximation remains accurate for longer times for small self-appraisal values.
We hypothesize that the inaccuracy of our mean-field approximation beyond short times may stem from our assumption that the opinions of the neighbors of each node
are representative of the global opinion distribution in a network.
As nodes update their opinions, connected subnetworks of nodes with similar opinions emerge, in violation of this homogeneity assumption.
It is worth investigating this hypothesis.

There are many ways to build on our work.
First, we do not know if our synchronously updating weighted-median {opinion} model is guaranteed to converge to a limit opinion distribution. 
There are known convergence criteria for bounded-confidence opinion models~\cite{Lorenz2010Convergence}, Mei et al.'s asynchronously updating weighted-median opinion model~\cite{mei2022convergence}, and other synchronously updating weighted-median opinion models~\cite{ZhangWMMPrejudice2025}.
However, these existing results do not apply directly to our model, as they rely either on the opinion-update rule being a contraction mapping on opinion space or on the opinion space consisting of a finite number of states.
Neither of these situations holds in our weighted-median opinion model.

In our weighted-median opinion model, we assumed that self-appraisal is constant with time and that it is homogeneous across all individuals.
One can relax both of these assumptions in a way that is reminiscent of the DeGroot--Friedkin model~\cite{Jia2015}.
Moreover, our mean-field approximation does not account for asymmetric relationships, and it is desirable to develop approximations that allow one to investigate such situations.
One can also study how media and other sources of information affect node opinions \cite{BrooksIdeologyContent2020} and extend our model to cover multidimensional and interrelated opinions \cite{MedoFragility2021}. 

Our work sheds light on opinion dynamics with weighted-median mechanisms of opinion updates~\cite{MeiMicro-Foundation2022, mei2022convergence, Han2024continuoustime, LiWeightedMedian2022, ZhangWMMPrejudice2025}.
Opinion models with opinion updates that are based on medians offer an interesting complement to the much more common mean-based models, and it is worthwhile to explore them further.
It is particularly relevant to evaluate which real-world opinion-evolution settings are better described by median-based opinion models and which are better described by mean-based opinion updates.


\appendix


\section{Final opinion distributions of our weighted-median model (\ref{eq:WMMI}, \ref{Theory:def:weighted_median})}
\label{sec:limit_opinion_app}

\subsection{Effect of self-appraisal on the final opinion distribution}\label{App:sec:limit_dist}

In Figures \ref{res:fig:Grid_WMMI_limit_dist}--\ref{res:fig:Georgetown_WMMI_limit_dist}, we show the final opinion distributions of our weighted-median opinion model {(\ref{eq:WMMI}, \ref{Theory:def:weighted_median})} on a square-lattice network, the WS network, and the Georgetown Facebook friendship network for several self-appraisal values and several different initial opinion distributions (see \cref{Sec:initial_sim_2}). 
The synthetic networks each have $2500$ nodes, and the Georgetown network has $9388$ nodes. 
We observe that the final opinion distributions are qualitatively similar for the different self-appraisal values.

\begin{figure}[h!]
    \centering
    \includegraphics[width = 0.9 \textwidth]{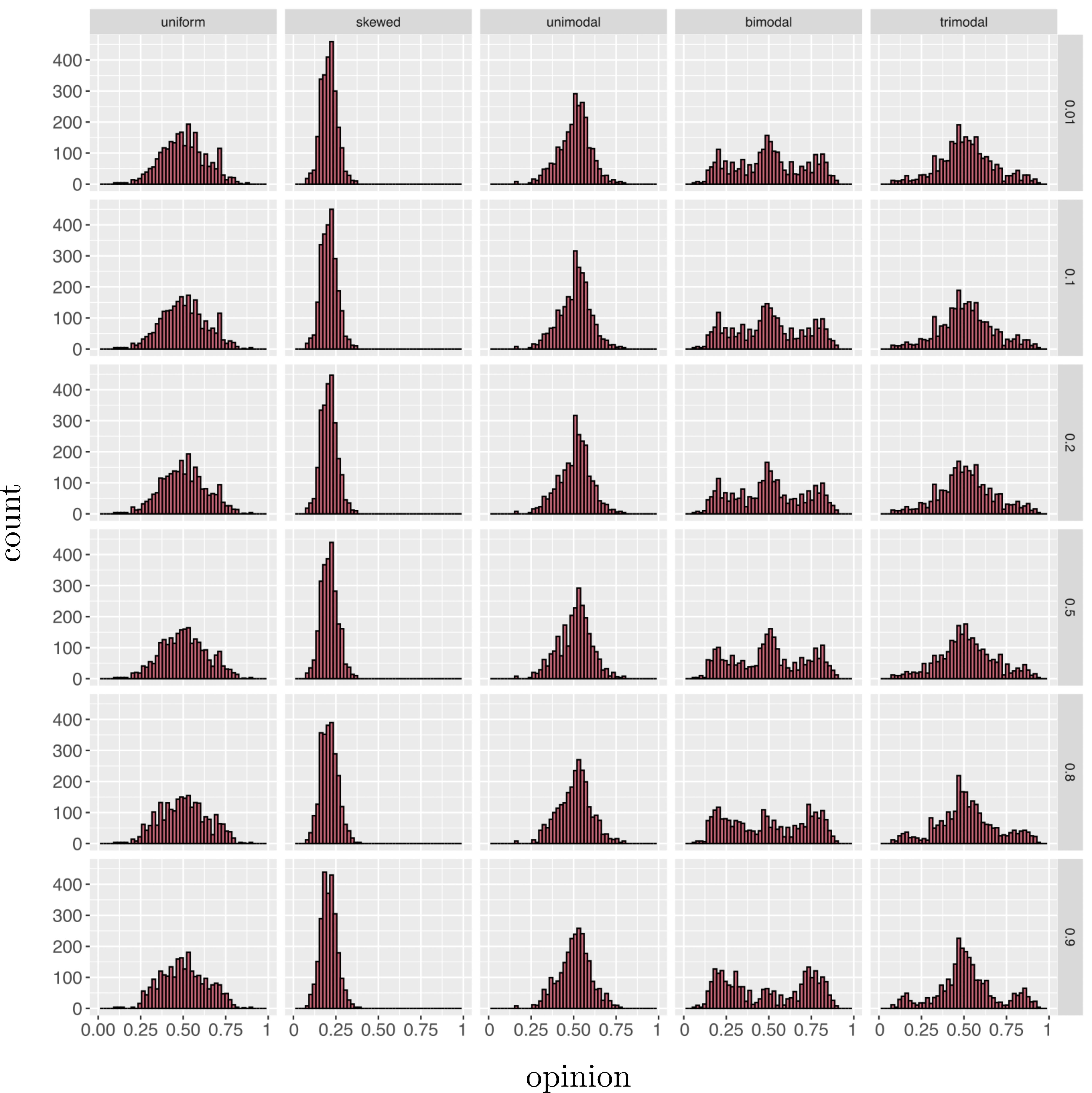}
    \caption{
        Final opinion distributions of our weighted-median {opinion} model {(\ref{eq:WMMI}, \ref{Theory:def:weighted_median})} for a square-lattice network for several initial opinion distributions and self-appraisal values.
    Each column has the same initial opinion distribution, and each row has the same self-appraisal value.
    The final opinion distributions in each column are qualitatively similar to each other, so it seems that the self-appraisal value has 
    little effect on the final opinion distribution for the square-lattice network.
    }
    \label{res:fig:Grid_WMMI_limit_dist}
\end{figure}

\begin{figure}[h!]
    \centering
    \includegraphics[width = 0.95 \textwidth]{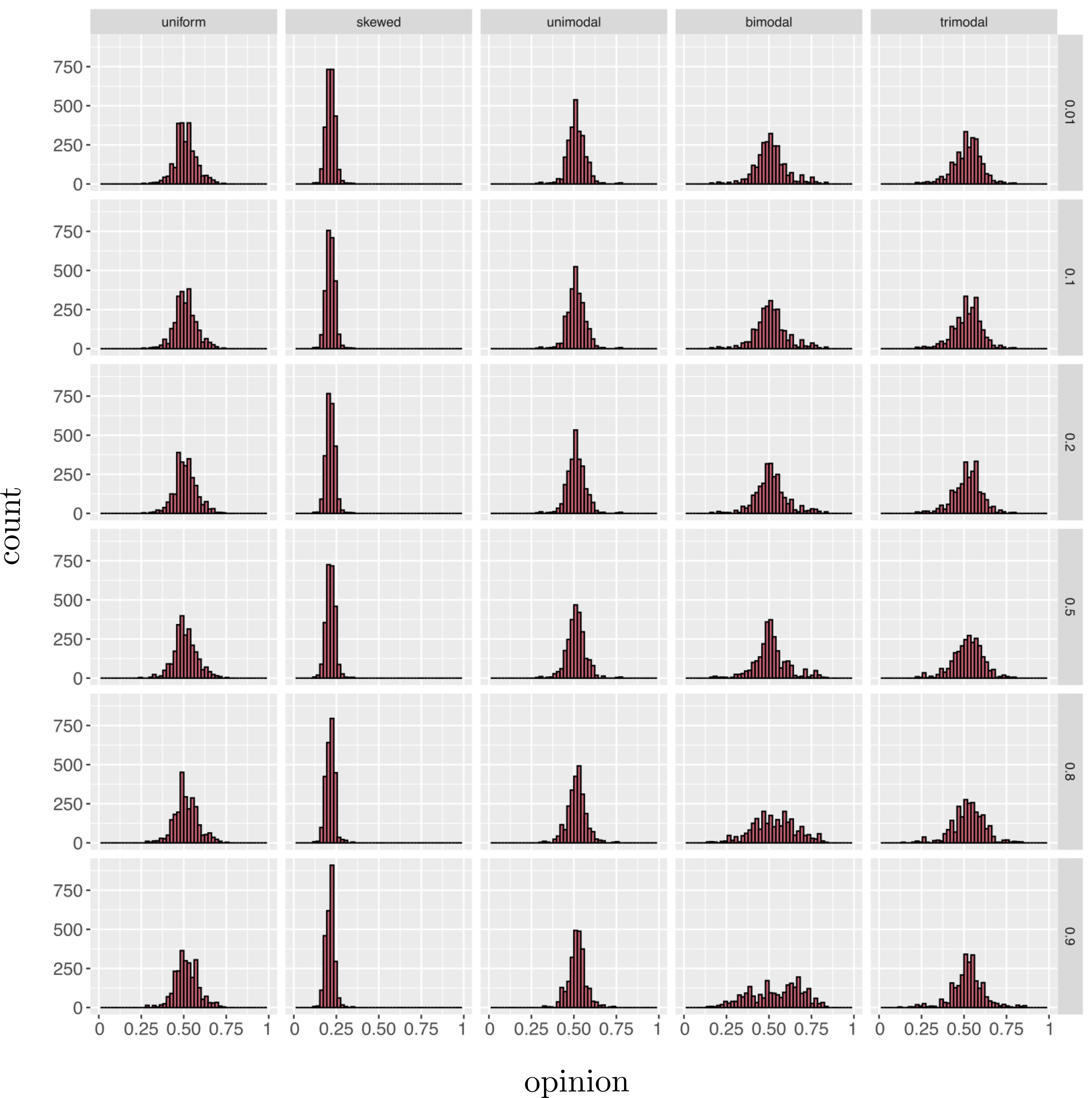}
    \caption{
        Final opinion distributions of our weighted-median {opinion} model {(\ref{eq:WMMI}, \ref{Theory:def:weighted_median})} for our WS network for several initial opinion distributions and self-appraisal values.
    Each column has the same initial opinion distribution, and each row has the same self-appraisal value.
    The final opinion distributions in each column are qualitatively similar to each other, so it seems that the self-appraisal value has little effect on the final opinion distribution for our WS network.
    }
    \label{res:fig:Watts-Strogatz_WMMI_limit_dist}
\end{figure}

\begin{figure}[h!]
    \centering
    \includegraphics[width = 0.95 \textwidth]{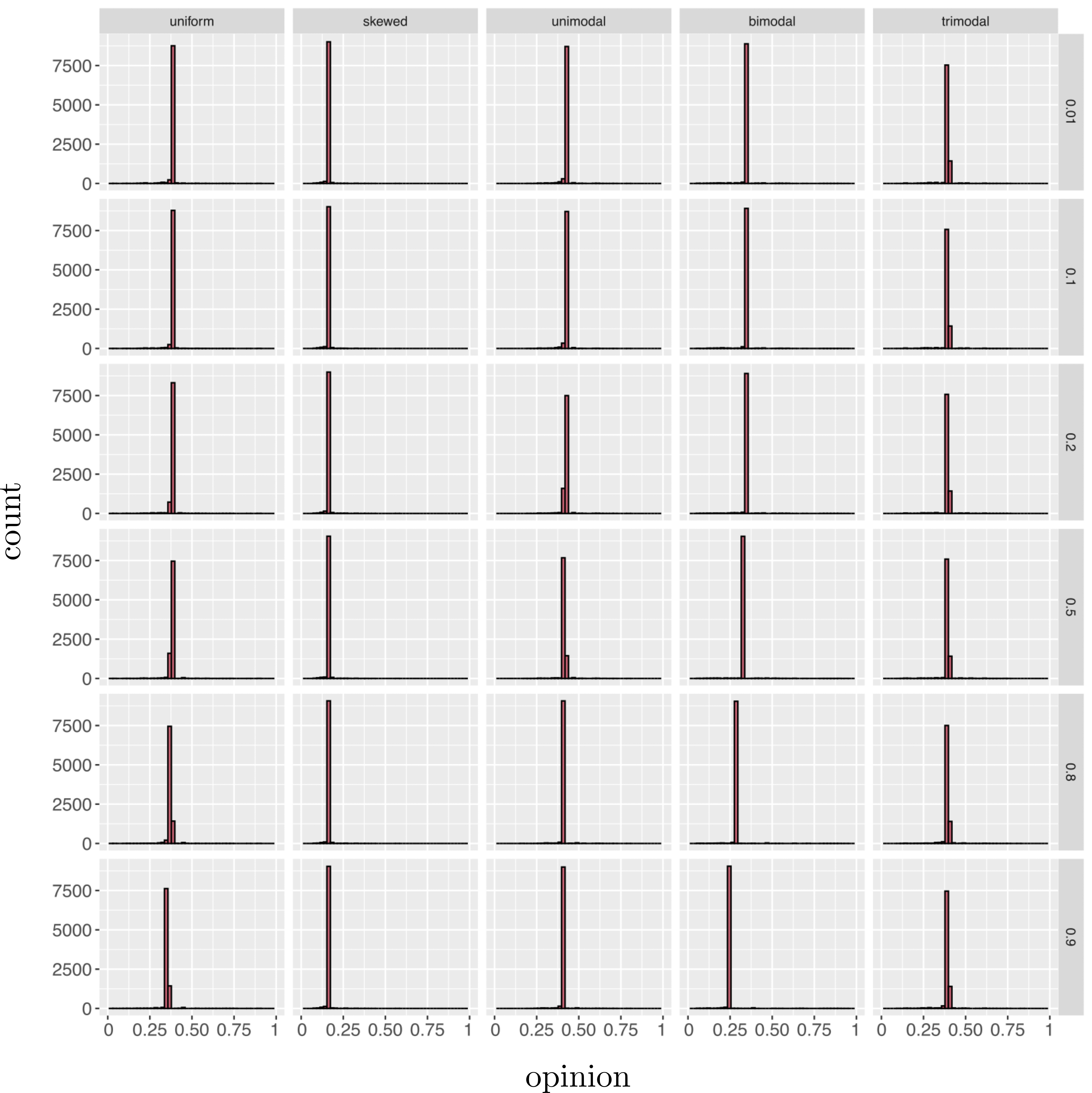}
    \caption{Final opinion distributions of our weighted-median {opinion} model {(\ref{eq:WMMI}, \ref{Theory:def:weighted_median})} for the Georgetown Facebook friendship network for several initial opinion distributions and self-appraisal values.
    Each column has the same initial opinion distribution, and each row has the same self-appraisal value.
    The final opinion distributions in each column are qualitatively similar to each other, so it seems that the self-appraisal value has little effect on the final opinion distribution for the Georgetown Facebook friendship network.
    }
    \label{res:fig:Georgetown_WMMI_limit_dist}
\end{figure}


\subsection{Summary statistics for the opinion-cluster sizes of the final opinion distributions}\label{App:sum_stat_lim_dist}

In Tables \ref{Res:tab:Lim_dist_group_stat_cycle}--\ref{Res:tab:Lim_dist_group_stat_obamacare}, we give summary statistics for the opinion-cluster sizes of the final opinion distributions of our weighted-median {opinion} model (\ref{eq:WMMI}, \ref{Theory:def:weighted_median}) for the networks in \cref{Sec:initial_sim_1}.

\begin{table}[]
    \centering
        \begin{tabular}{l|lllll}
        Statistic   & Uniform  & Unimodal & Skewed& Bimodal & Trimodal \\ \hline
        Number of opinion clusters & 1019 & 986 & 999 & 992 & 988 \\
        Mean size & 2.51 & 2.48 & 2.54 & 2.53 & 2.54 \\
        Variance of the size  & 0.45 & 0.53 & 0.50 & 0.54 & 0.53 \\
        Kurtosis of the size   & 0.90  & 1.09 & 1.05  & 1.16 & 1.09
        \end{tabular}
    \caption{
        Summary statistics for the opinion clusters of the final opinion distribution of our weighted-median opinion model with self-appraisal $s = 0.7$ on a 2500-node cycle network for several initial opinion distributions.
    For each initial opinion distribution, we perform a single simulation until the convergence criterion \cref{sim:conv_critera} is satisfied.
    }
    \label{Res:tab:Lim_dist_group_stat_cycle}
\end{table}

\begin{table}[h!]
    \centering
        \begin{tabular}{l|lllll}
        Statistic  & Uniform  & Unimodal & Skewed & Bimodal & Trimodal \\ \hline
        Number of opinion clusters & 925 & 941 & 946 & 917 & 919    \\
        Mean size  & 2.70  & 2.66 & 2.64 & 2.72 & 2.72 \\
        Variance of the size           & 0.98  & 1.00 & 0.97  & 1.08 & 1.05 \\
        Kurtosis of the size           & 2.12 & 2.78 & 2.43  & 3.29   & 2.39
        \end{tabular}
    \caption{
        Summary statistics for the opinion clusters of the final opinion distribution of our weighted-median opinion model with self-appraisal $s = 0.7$ on the 2500-node prism network for several initial opinion distributions.
    For each initial opinion distribution, we perform a single simulation until the convergence criterion \cref{sim:conv_critera} is satisfied.
    }
    \label{Res:tab:Lim_dist_group_stat_prism}
\end{table}

\begin{table}[h!]
    \centering
        \begin{tabular}{l|lllll}
        Statistic                   & Uniform & Unimodal & Skewed & Bimodal & Trimodal \\ \hline
        Number of opinion clusters  & 783     & 785      & 779    & 800     & 792      \\
        Mean size                   & 3.19    & 3.18     &3.21    & 3.13   & 3.16    \\
        Variance of the size        & 2.49    & 2.44     & 2.48   & 1.91   & 2.44 \\
        Kurtosis of the size        & 55.15   & 46.74    & 45.03  & 26.59 & 52.92
        \end{tabular}
    \caption{
        Summary statistics for the opinion clusters of the final opinion distribution of our weighted-median opinion model with self-appraisal $s = 0.7$ on the 2500-node square-lattice network for several initial opinion distributions.
    For each initial opinion distribution, we perform a single simulation until the convergence criterion \cref{sim:conv_critera} is satisfied.
    }
    \label{Res:tab:Lim_dist_group_stat_grid}
\end{table}

\begin{table}[]
    \centering
        \begin{tabular}{l|lllll}
        Statistic                   & Uniform               & Unimodal              & Skewed               & Bimodal & Trimodal \\ \hline
        Number of opinion clusters  & 395                   & 393                   & 378                  & 375 &415 \\
        Mean size                   & 6.32                  & 6.36                  & 6.61                 & 6.67 &6.02 \\
        Variance of the size        & 25.57                 & 24.95                 & 27.12                & 24.73 & 20.57 \\
        Kurtosis of the size        & $1.32 \times 10^{4}$  & $1.52 \times 10^{4}$  & $1.21 \times 10^{4}$ & $5.70 \times 10^{3}$ & $1.14 \times 10^{4}$
        \end{tabular}
    \caption{
        Summary statistics for the opinion clusters of the final opinion distribution of our weighted-median opinion model with self-appraisal $s = 0.7$ on a 2500-node WS network for several initial opinion distributions.
    For each initial opinion distribution, we perform a single simulation until the convergence criterion \cref{sim:conv_critera} is satisfied.
    }
    \label{Res:tab:Lim_dist_group_stat_WS}
\end{table}

\begin{table}[]
    \centering
        \begin{tabular}{l|lllll}
        Statistic                   & Uniform                 & Unimodal               & Skewed                 & Bimodal   & Trimodal  \\ \hline
        Number of opinion clusters  & 43                      & 40                     & 57                     & 41 & 47   \\
        Mean size                   & 58.14                   & 62.50                  & 43.86                  & 60.98                  & 53.19 \\
        Variance of the size        & $8.09 \times 10^{4}$  & $8.88 \times 10^{4}$ & $6.32 \times 10^{4}$ & $8.57 \times 10^{4}$  & $8.90 \times 10^{4}$ \\
        Kurtosis of the size        & $2.63 \times 10^{11}$   & $2.93 \times 10^{11}$  & $2.17 \times 10^{11}$  & $2.81 \times 10^{11}$   & $3.54 \times 10^{11}$
        \end{tabular}
    \caption{
        Summary statistics for the opinion clusters of the final opinion distribution of our weighted-median opinion model with self-appraisal $s = 0.7$ on a 2500-node BA network for several initial opinion distributions.
    For each initial opinion distribution, we perform a single simulation until the convergence criterion \cref{sim:conv_critera} is satisfied.
    }
    \label{Res:tab:Lim_dist_group_stat_BA}
\end{table}

\begin{table}[h!]
    \centering
        \begin{tabular}{l|lllll}
        Statistic                  & Uniform              & Unimodal             & Skewed               & Bimodal & Trimodal  \\ \hline
        Number of opinion clusters & 90                   & 58                   & 210                  & 266     & 78 \\
        Mean size                  & 8.47                 & 13.14                & 3.6285               & 2.86    & 9.77 \\
        Variance of the size       & 3685.45              & 5117.64              & 29.05                & 11.05   & 3612.15 \\
        Kurtosis of the size       & $1.19 \times 10^{9}$ & $1.44 \times 10^{9}$ & $3.15 \times 10^{4}$ & 2048.04 & $9.11 \times 10^{8}$
        \end{tabular}
    \caption{
        Summary statistics for the opinion clusters of the final opinion distributions of our weighted-median opinion model with self-appraisal $s = 0.7$ on the Caltech Facebook friendship network for several initial opinion distributions.
    For each initial opinion distribution, we perform a single simulation until the convergence criterion \cref{sim:conv_critera} is satisfied.
    }
    \label{Res:tab:Lim_dist_group_stat_Caltech}
\end{table}

\begin{table}[h!]
    \centering
        \begin{tabular}{l|lllll}
        Statistic                  & Uniform               & Unimodal & Skewed                & Bimodal   & Trimodal  \\ \hline
        Number of opinion clusters & 289                   & 1417     & 248                   & 332                  & 1185 \\
        Mean size                  & 7.7854                & 1.5878   & 9.574                 & 6.7771               & 1.8987 \\
        Variance of the size       & 6684.79               & 2.3467   & 7694.07               & 1253.13              & 79.14\\
        Kurtosis of the size       & $1.27 \times 10^{10}$ & 235.37   & $1.29 \times 10^{10}$ & $2.55 \times 10^{7}$ & $5.40 \times 10^{6}$
        \end{tabular}
    \caption{
        Summary statistics for the opinion clusters of the final opinion distributions of our weighted-median opinion model with self-appraisal $s = 0.7$ on the Bowdoin Facebook friendship network for several initial opinion distributions.
    For each initial opinion distribution, we perform a single simulation until the convergence criterion \cref{sim:conv_critera} is satisfied.
    }
    \label{Res:tab:Lim_dist_group_stat_Bowdoin}
\end{table}

\begin{table}[]
    \centering
        \begin{tabular}{l|lllll}
        Statistic                  & Uniform              & Unimodal             & Skewed                & Bimodal               & Trimodal  \\ \hline
        Number of opinion clusters & 2516                 & 3686                 & 3629                  & 5303                  & 4151 \\
        Mean size                  & 3.73                 & 2.55                 & 2.59                  & 1.77                  & 2.26 \\
        Variance of the size       & 1365.86              & 201.77               & 101.26                & 287.42                & 497.17 \\
        Kurtosis of the size       & $2.20 \times 10^{9}$ & $6.24 \times 10^{8}$ & $1.43 \times 10^{9}$  & $3.83 \times 10^{8}$  & $6.07 \times 10^{8}$
        \end{tabular}
    \caption{
        Summary statistics for the opinion clusters of the final opinion distribution of our weighted-median opinion model 
    with self-appraisal $s = 0.7$ on the Georgetown Facebook friendship network for several initial opinion distributions.
    For each initial opinion distribution, we perform a single simulation until the convergence criterion \cref{sim:conv_critera} is satisfied.
    }
    \label{Res:tab:Lim_dist_group_stat_Georgetown}
\end{table}

\begin{table}[]
    \centering
        \begin{tabular}{l|lllll}
        Statistic                  & Uniform               & Unimodal & Skewed                & Bimodal               & Trimodal \\ \hline
        Number of opinion clusters & 1417                  & 4414     & 3472                  & 1539                  & 2508 \\
        Mean size                  & 5.65                  & 2.81     & 2.31                  & 5.20                  & 3.19 \\
        Variance of the size       & $1.29 \times 10^{4}$  & 3.11     & 53.02                 & 3191.15               & 199.22 \\
        Kurtosis of the size       & $1.73 \times 10^{12}$ & 516.61   & $1.81 \times 10^{7}$  & $4.90 \times 10^{10}$ & $2.52 \times 10^{8}$ 
        \end{tabular}
    \caption{
        Summary statistics for the opinion clusters of the final opinion distribution of our weighted-median opinion model with self-appraisal $s = 0.7$ on the Obamacare Twitter followership network for several initial opinion distributions.
    For each initial opinion distribution, we perform a single simulation until the convergence criterion \cref{sim:conv_critera} is satisfied.
    }
    \label{Res:tab:Lim_dist_group_stat_obamacare}
\end{table}

\begin{table}[h!]
    \centering
        \begin{tabular}{l|lllll}
        Statistic                  & Uniform               & Unimodal               & Skewed                & Bimodal                & Trimodal \\ \hline
        Number of opinion clusters & 1123                  & 1007                   & 1044                  & 1025                   & 1077 \\
        Mean size                  & 5.44                  & 6.07                   &5.86                   & 5.96                   & 5.68 \\
        Variance of the size       & 5594.65               & $1.09 \times 10^{5}$ & 9837.41               & $1.26 \times 10^{4}$ & 9939.84  \\
        Kurtosis of the size       & $2.06 \times 10^{10}$ & $8.95 \times 10^{10}$  & $8.40 \times 10^{11}$ & $1.44 \times 10^{11}$  & $8.56 \times 10^{10}$ 
        \end{tabular}
    \caption{
        Summary statistics for the opinion clusters of the final opinion distributions of our weighted-median opinion model with self-appraisal $s = 0.7$ on the abortion Twitter followership network for several initial opinion distributions.
    For each initial opinion distribution, we perform a single simulation until the convergence criterion \cref{sim:conv_critera} is satisfied.
    }
    \label{Res:tab:Lim_dist_group_stat_abortion}
\end{table}



\section{ {Derivation, accuracy, and numerical approximations of the mean-field approximation (\ref{Theory:eq:mean_field_eq}, \ref{Theory:eq:global_dens})}}
\label{sec:mean_field_app}

In this appendix, we derive the mean-field approximation {(\ref{Theory:eq:mean_field_eq}, \ref{Theory:eq:global_dens})}, present the specifications of our tests of its accuracy, investigate inaccuracies from finite-size effects and annealing-assumption effects, and discuss some properties of and issues in numerical solutions of \cref{Theory:eq:mean_field_eq}.


\subsection{Derivation of the mean-field approximation {(\ref{Theory:eq:mean_field_eq}, \ref{Theory:eq:global_dens})}}\label{App:MF_derivation}

We develop a degree-based mean-field approximation of our weighted-median opinion model {(\ref{eq:WMMI}, \ref{Theory:def:weighted_median})} for configuration-model networks with a prescribed degree distribution \cite{Fosdick2016ConfiguringSequences}.
A degree-$k$ node has an associated opinion distribution $P_k(x,t)$ at time $t$.
With our mean-field assumptions, which are inspired by the ones in \cite{fennell2021}, we derive a coupled system of difference equations for the time evolution of $P_k(x,t)$.

In our derivation, we make two key assumptions. 
First, we make an annealed-network assumption by rewiring the edges of the configuration-model network at each time step while fixing the network's degree sequence~\cite{Dorogovtsev2008CriticalNetworks}. 
Therefore, the network at each time step is one instantiation of our configuration-model ensemble.
Second, we disregard the edge weights and edge directions.
Therefore, we use adjacency matrices instead of influence matrices.

Because we ignore edge weights and directions, at any time $t \in \mathbb{N}_0 = \{0, 1, \ldots \}$, the network $G = (V,E)$ (which we assume does not have any isolated nodes) has an associated unweighted and symmetric adjacency matrix $A$, so $A_{ij} = 1$ if $(i,j) \in E$ and $A_{ij} = 0$ otherwise.

Disregarding edge weights necessitates defining an unweighted analogue of the weighted-median opinion model (\ref{eq:WMMI}, \ref{Theory:def:weighted_median}).
Let {$A$ be a symmetric adjacency matrix,} $n(i)$ denote the number of neighbors of node $i$, the scalar $x_i(t) \in [0,1]$ denote the opinion of node $i \in V$ at time $t \in \mathbb{N}_0$, and the state $x(t) = (x_1(t), x_2(t), \ldots, x_N(t))$ denote the vector of the $N = |V|$ node opinions.
Given an initial state $x(0)$, our {median} opinion model is the map
\begin{equation} \label{map-app}
    x_i(t + 1) = (1 - s)x_i(t) + s\, \mathrm{Med}_i(x(t);A)\,, \quad i \in V\,,
\end{equation}
    where $s \in (0,1)$ is the amount of self-appraisal and $\mathrm{Med}_i(x(t);A)$ denotes the {median} of the opinion states $x(t)$ with respect to the adjacency-matrix entries $A_{i1}, \, A_{i2},\, A_{i3}, \, \ldots\,, A_{iN}$.
    More precisely, 
    $\mathrm{Med}_i(x(t);A) = x^{*}\in \mathbb{R}$ if $x^{*}$ satisfies
    \begin{equation}\label{App:def:median}
        \sum_{ \{j\, : \, x_j <  x^{*} \}} A_{{i}j} \, {\leq}\, \frac{1}{2}n(i) \quad \text{and} \quad
        \sum_{\{j\, : \, x_j > x^{*}\}} A_{{i}j}\, {\leq}\, \frac{1}{2}n(i)\,.
    \end{equation}
 If multiple values of $x^{*}$ satisfy equation \cref{App:def:median}, we let $\mathrm{Med}_i(x(t);A)$ be the value that is closest to $x_i(t)$. If two values of $\mathrm{Med}_i(x(t);A)$ are equally close to $x_i(t)$, then we select the smaller of those two values.

If $G$ has multiple components, then node opinions evolve independently on each component. 
Therefore, we assume for convenience that $G$ is connected.
One can view the weighted-median {opinion} model {(\ref{eq:WMMI}, \ref{Theory:def:weighted_median})} of the main manuscript as a generalization of the {median opinion} model (\ref{map-app}, \ref{App:def:median}) because one can also consider weighted edges in the earlier model.

Let $q_k$ denote the probability that a node that we select uniformly at random from the set $V$ has degree $k \in \mathbb{N}_0$.
By assumption, there are no isolated nodes, so $q_0 = 0$.
Let $D$ denote the set of distinct degrees of the nodes.
For each $k$, let $P_k(x, t)$ denote the distribution of the opinions of degree-$k$ nodes at time $t$.
The probability that a degree-$k$ node's opinion is in the interval $[x,  x + \Delta x)$ at time $t \in \mathbb{N}_0$ is thus $P_k(x,t)\Delta x + \mathcal{O}((\Delta x)^2)$.
 
We derive a system of coupled finite-difference equations for the evolution of the densities $P_k(x, t)$ for $k \in D$.
For large $N$ and small $\Delta x > 0$, the expected number of degree-$k$ nodes with opinions in the interval $[x,  x + \Delta x)$ is $N q_k P_k(x,t) \Delta x$ at time $t$.
The expected change of this number from time $t$ to time $t + 1$ is
\begin{align}\label{Theory:eq:mean_field_1}
    Nq_kP_k(x,t + 1) \Delta x - Nq_k P_k(x,t) \Delta x  \,.
\end{align}

To obtain an expression for $P_k(x, t + 1)$ as a function of $P_k(x,t)$, we use equation \cref{Theory:eq:mean_field_1} and express the expected change in the number of degree-$k$ nodes with opinions in the interval $[x,  x + \Delta x)$ from time $t$ to time $t + 1$ in terms of $P_k(x, t)$. The latter expression is $N q_k P_k(x,t) \Delta x$, which we obtain as follows.
First, we derive the distribution of the opinion of a single neighbor of a degree-$k$ node.
We then derive the distribution of the weighted-median opinion of the neighbors of a degree-$k$ node.
Using the latter quantity, we derive the mean number of opinions of degree-$k$ nodes that enter the interval $[x,  x + \Delta x)$ at time $t + 1$ and the mean number of opinions of degree-$k$ nodes that leave the interval $[x,  x + \Delta x)$ at time $t + 1$.

\subsubsection{Opinion distribution of a uniformly random neighboring node}

We now derive the opinion distribution of a uniformly random neighbor of a degree-$k$ node.
Because the opinions evolve on a configuration-model network, the probability that a given edge that is incident to a degree-$k$ node is incident at its other end
to a specific degree-$l$ node is $\frac{l}{2|E| - 1}$.
As $N \rightarrow \infty$, the expected number of degree-$l$ nodes is approximately $2\frac{q_l |E|}{\sum_{i \in D} i q_i}$, where $\sum_{i \in D} i q_i$ is the mean degree.
Therefore, as $N \rightarrow \infty$, the probability that a given edge that is incident to a degree-$k$ node is also incident to a degree-$l$ node is
\begin{align}\label{Theory:eq:mean_field_2}
    \pi_{l} = \frac{l q_l}{{\sum_{i \in D} i q_i}}\,.
\end{align}

We now derive the opinion distribution of a uniformly random neighbor of a node.
Because we determine the edges uniformly at random at each time step, this opinion distribution is independent of that node's degree in the limit $N \rightarrow \infty$.
The probability that a uniformly random neighbor has an opinion in the interval $[x,  x + \Delta x)$ is 
\begin{align}\label{Theory:eq:mean_field_3}
    &\mathbb{P}(x_j(t) \in [x,  x + \Delta x) \, \cap \, e = (i,j)\in E\: |\: i,\, j \in V )\nonumber\\
    &\quad\, = \sum_{l\in D}\mathbb{P}(\text{there is an edge between node $i$ and a degree-$l$ node})\, P_l(x,t)\Delta x\\
    &\quad\, = \sum_{l\in D}\pi_{l} P_l(x,t)\Delta x\nonumber\\
    &\quad\, = \phi(x,t) \Delta x \, ,\nonumber
\end{align}
where $\phi(x,t) = \sum_{l\in D}\pi_{l} P_l(x,t)$ is the opinion distribution of a uniformly random neighbor of a node in the limit $N \rightarrow \infty$ and $D$ is the set of distinct degrees. Let $\Phi(x,t) = \int_{0}^x \phi(u,t) \, \mathrm{d} u$ denote the associated cumulative distribution function.


\subsubsection{Distribution of the weighted-median opinion of neighboring nodes}\label{sec:MF_theta_der}
Using \cref{Theory:eq:mean_field_3}, we derive the distribution of the median opinion of the neighbors of a degree-$k$ node.
To do this, we introduce some notation and results for order statistics.


\paragraph{Order statistics}


Suppose that $X_{1}, X_2, \ldots , X_k$ are independent univariate random variables with a common continuous probability density function $f_X$ and associated cumulative distribution function $F_X$.
We order these random variables from smallest to largest. The $r$th order statistic $X_{(r)}$ describes the $r$th-smallest value of a quantity and has the probability density function \cite{Sarndal2004OrderStatistics}
\begin{align}\label{Theory:eq:order_stat}
    f_{X_{(r)}}(x) = \frac{n  !}{(r - 1)!(n - r)!} f_X(x) \left(F_X(x)\right)^{r - 1} \left(1 - F_x(x)\right)^{n - r} \, .
\end{align}
The conditional distribution of the $(r+1)$th order statistic given the $r$th order statistic is 
\begin{align} \label{Theory:eq:mean_field_14}
    f_{\{X_{(r + 1)}  | X_{(r)} = x\}}\,(y) = (n - r)\frac{f_X(y)}{1 - F_X(x)} \left(\frac{1 - F(y)}{1 - F(x)}\right)^{n - r - 1}.
\end{align}


\paragraph{Distribution of the weighted-median opinion of neighboring nodes}


We now derive the distribution of the median opinion of neighboring nodes in a configuration-model network.
Let node $i$ have degree $k$.
We use separate arguments for odd and even $k$. 

First, we consider odd $k$. Let $k = 2m + 1$.
Because the opinions of the neighbors of node $i$ follow the common distribution $\phi(x,t)$ in the limit $N \rightarrow \infty$, we view them as independent and identically distributed random variables.
The median opinion $\mathrm{Med}_i(x,A)$ is the $(m + 1)$th order statistic.
It has the distribution 
\begin{align}
    \phi_{x_{(m + 1)_i}}(x,t) = \frac{(2m + 1)!}{m!m!} \phi_{2m + 1}(x,t) \Phi_{2m + 1}(x,t)^m \left(1 - \Phi(x,t)\right)^m\,.
\end{align}
Let $\theta_{2m + 1}(x,t)$ be the distribution of the median neighbor opinion of a node with degree $k = 2m + 1$.
We can then write
\begin{align}\label{Theory:eq:mean_field_15}
    \theta_{2m + 1}(x,t) = \frac{(2m + 1)!}{m!m!} \phi_{2m + 1}(x,t) \Phi_{2m + 1}(x,t)^m \left(1 - \Phi(x,t)\right)^m\,.
\end{align}

We now consider even $k$.
Let $k = 2m$.
The quantity $\mathrm{Med}_i(x(t);A)$ is either the $m$th or $(m+1)$th order statistic, depending on which of them is closer to $x_i(t)$.
Given the definition of the {median in \cref{App:def:median}}, we choose $\mathrm{Med}_i(x(t);A)$ to be the $m$th order statistic if they are equally close.
Let $x_{(m)_i}$ and $x_{(m + 1)_i}$, respectively, be the $m$th and $(m + 1)$th order statistics of the opinions of the $2m$
neighbors of node $i$.
In the limit $N \rightarrow \infty$, the neighbor opinions follow a common distribution $\phi(x,t)$.
We derive an expression for the probability
 \begin{align}\label{Theory:eq:mean_field_13}
    \mathbb{P}(\mathrm{Med}_i(x(t);A) = x_{(m + 1)_i})&=
    \mathbb{P}\left( \, |x_i - x_{(m + 1)_i}| < |x_i - x_{(m)_i}| \right)\, ,
\end{align}
where the opinion $x_i$ follows the distribution $P_{2m}(x,t)$.

First suppose that $x_i < x_{(m)_i}$.
In this case, the probability \cref{Theory:eq:mean_field_13} is $0$ because the $(m + 1)$th order statistic is larger than $x_{(m)_i}$.

Now suppose that $x_i \geq x_{(m)_i}$ and let $c = x_i - x_{(m)_i}$. 
Equation \cref{Theory:eq:mean_field_13} gives the probability that the $(m + 1)$th order statistic is in the interval $[x_{(m)_i}, x_{(m)_i} + 2c)$.
Using the law of total probability \cite{Pitman1993Probability}, we then write
\begin{align}
    &\mathbb{P}(\mathrm{Med}_i(x(t);A) = x_{(m+1)_i}) = \\
    & \hspace{0.5cm}\int_{x_i\in [0,1]}\int_{x_{(m)_i}\leq x_i} \int_{y\in [x_{(m)_i}, -x_{(m)_i} + 2x_i)}  \phi_{\{x_{(m + 1)_i}  | x_{(m)_i}\}}(y,t)\, P_k(x_i,t)\, \phi_{x_{(m)_i}}(x_{(m)_i},t) \, \mathrm{d} y  \, \mathrm{d} x_{(m)_i}\, \mathrm{d} x_i\, , \nonumber
\end{align}
where $\phi_{\{x_{(m+1)_i}  | x_{(m)_i}\}}$ is the distribution of the $(m + 1)$th order statistic conditioned on the $m$th order statistic and we note that $[x_{(m)_i}, x_{(m)_i} + 2c) = [x_{(m)_i},  -x_{(m)_i} + 2x_i)$.
From equations \cref{Theory:eq:order_stat} and \cref{Theory:eq:mean_field_14}, we see that
\begin{align*}
    \phi_{\{x_{(m+1)_i}  | x_{(m)_i}\}}(y,t) &= m\frac{\phi(y)}{1 - \Phi(x_{(m)_i},t)}\left(\frac{1 - \Phi(y,t)}{1 - \Phi(x_{(m)_i},t)}\right)^{m-1} \,, \\
    \phi_{x_{(m)_i}}(b,t)  &= \frac{(2m)!}{(m-1)!m!} \phi(b,t) \left( \Phi(b,t)\right)^{m-1}\left( 1- \Phi(b,t)\right)^m
\end{align*}
for $y\geq b$.

Therefore,
\begin{align}
    \mathbb{P}(\mathrm{Med}_i(x(t);A) = x_{(m+1)_i}) 
     &= 
    \int_{x_i\in [0,1]}\int_{x_{(m)_i}\leq x_i} \int_{y\in [x_{(m)_i}, -x_{(m)_i} + 2x_i)} 
    m\frac{\phi(y,t)}{1 - \Phi(x_{(m)_i},t)}\left(\frac{1 - \Phi(y,t)}{1 - \Phi(x_{(m)_i},t)}\right)^{m-1} \nonumber\\
    &\qquad\qquad\quad\quad\quad\quad\quad\quad \times 
    \frac{(2m)!}{(m - 1)!m!} \phi(x_{(m)_i},t) \left( \Phi(x_{(m)_i},t)\right)^{m - 1}\left( 1- \Phi(x_{(m)_i},t)\right)^m \nonumber\\
    &\qquad\qquad\quad\quad\quad\quad\quad\quad \times P_{2m}(x_i,t) \,\mathrm{d} y  \, \mathrm{d} x_{(m)_i}\, \mathrm{d} a\nonumber\\
    &=    
    \frac{(2m)!}{((m - 1)!)^2}\int_{x_i\in [0,1]}  P_{2m}(x_i,t) \int_{x_{(m)_i}\leq x_i} \phi(x_{(m)_i},t) \left( \Phi(x_{(m)_i},t)\right)^{m-1} \nonumber \\
    &\qquad\qquad\quad\quad\quad\quad\quad\quad \times \int_{y\in [x_{(m)_i}, -x_{(m)_i} + 2x_i)} \phi(y,t)\left(1 - \Phi(y,t)\right)^{m - 1} \,\mathrm{d} y  \, \mathrm{d} x_{(m)_i}\, \mathrm{d} x_i \, .\nonumber
\end{align}

We define the notation $p_{2m}(t) = \mathbb{P}(\mathrm{Med}_i(x(t);A) = x_{(m + 1)_i})$, and we express $\theta_{2m}$ using the $m$th and $(m + 1)$th order statistics of $\phi(x,t)$.
With $k = 2m$, we have
\begin{align}\label{Theory:eq:mean_field_16}
    \theta_{2m}(x,t) & = (1 - p_{2m}(t)) \phi_{x_{(m)}}(x,t) +  p_{2m}(t)\phi_{x_{(m + 1)}}(x,t) \nonumber \\
                     &= (1 - p_{2m}(t)) \frac{(2m)!}{(m - 1)!m!} \phi(x,t) \Phi(x,t)^{m - 1} \left(1 - \Phi(x,t)\right)^m \nonumber \\
                     &\qquad + p_{2m}(t)\frac{(2m)!}{(m - 1)!m!} \phi(x,t) \Phi(x,t)^{m} \left(1 - \Phi(x,t)\right)^{m - 1} \nonumber \\
    &= \frac{(2m)!}{(m - 1)!m!}\phi(x,t)\big(\Phi(x,t) \left(1 - \Phi(x,t)\right)\big)^{m - 1} \\
    &\hspace{2cm} \times \big((1 - p_{2m}(t))\left(1 - \Phi(x,t)\right) + p_{2m}(t) \Phi(x,t)\big)\,. 
    \nonumber 
\end{align}

From equations \cref{Theory:eq:mean_field_15} and \cref{Theory:eq:mean_field_16}, we see that the median neighbor opinion of a degree-$k$ node has the distribution 
\begin{align}\label{Theory:eq:mean_field_5}
        \theta_k(x,t) = 
            \begin{cases}
                \frac{(2m + 1)!}{m!m!} \phi_{2m + 1}(x,t) \Phi_{2m + 1}(x,t)^m \left(1 - \Phi(x,t)\right)^m\,, \quad \text{$k = 2m + 1$}
                \\[1.2em]
        				\begin{array}{l}
        					\hspace{-2mm}\frac{(2m)!}{(m - 1)!m!}\phi(x,t)\big(\Phi(x,t) \left(1 - \Phi(x,t)\right)\big)^{m - 1}\\
                            \hspace{0.5cm}\times \big((1 - p_{2m}(t))\left(1 - \Phi(x,t)\right) + p_{2m}(t) \Phi(x,t)\big)
        				\end{array}\,,
        	    \quad \text{$k = 2m$}\,,
            \end{cases} 
\end{align}
where
\begin{align}
    p_{2m}(t) &= \frac{(2m)!}{((m - 1)!)^2}\int_{a\in [0,1]}  P_{2m}(x_i,t) \int_{b\leq x_i} \phi(b,t) \left( \Phi(b,t)\right)^{m - 1}   \nonumber  \\
		&\quad\quad\quad\quad\quad\quad\quad\quad\quad\quad\quad \times \int_{y\in [b, -b + 2x_i)} \phi(y,t)\left(1 - \Phi(y,t)\right)^{m - 1} \,\mathrm{d} y  \,\mathrm{d} b\, \mathrm{d} x_i\,.
\end{align}


\subsubsection{Mean number of degree-\texorpdfstring{$k$}{TEXT} nodes with opinions that enter the interval \texorpdfstring{$[x, x + \Delta x)$}{TEXT} at time \texorpdfstring{$t + 1$}{TEXT}}\label{App:sec:mean_enter_interval}
Using \cref{Theory:eq:mean_field_5}, we derive the expected number of nodes whose opinions are outside the interval $[x,  x + \Delta x)$ at time $t$ but inside this interval at time $t + 1$.
Let $i \in V$ be a degree-$k$ node, and suppose that $x_i(t) = y \notin [x,  x + \Delta x)$.
For the opinion of node $i$ to enter $[x,  x + \Delta x)$, we need
\begin{align}\label{Theory:eq:mean_field_6}
    x \leq (1 - s)y + s \, \mathrm{Med}_i( x(t), A) < x + \Delta x \,,
\end{align}
which yields
\begin{align}
    \frac{x + (s - 1)y}{s} \leq  \mathrm{Med}_i & (x(t), A) <  \frac{x + (s - 1)y}{s} + \frac{\Delta x}{s} 
\end{align}
for $s \neq 0$.

Because node $i$ has degree $k$, we know that $\mathrm{Med}_i(x(t);A)$ has the distribution $\theta_k(x,t)$ in the limit $N \rightarrow \infty$.
The probability that \cref{Theory:eq:mean_field_6} holds is thus
\begin{align}
    \theta_k\left( \frac{x + (s - 1)y}{s}, t\right)\frac{\Delta x}{s}\,.
\end{align}

The expected number of degree-$k$ nodes that at time $t$ have an opinion in $[y,  y + \mathrm{d} y)$ but outside $[x, x + \Delta x)$ and then at time $t + 1$ have an opinion in $[x, x + \Delta x)$ is approximately
\begin{align}
    \frac{N q_k}{s} \theta_k\left( \frac{x + (s - 1)y}{s}, t\right)\frac{\Delta x}{s}P_k(y,t) \Delta x\, \mathrm{d} y \,.
\end{align}
The approximate nature of our statement arises because the intervals $[x, x+ \Delta x) $ and $[y, y + \mathrm{d} y)$ may overlap. However, this overlap disappears as $\Delta x \rightarrow 0$ and $\mathrm{d} y \rightarrow 0$.
Integrating over $y$, we see that the mean number of degree-$k$ nodes whose opinions enter $[x,  x + \Delta x)$ at time $t + 1$ is
\begin{align}\label{Theory:eq:mean_field_8}
    N q_k \int_{y\in [0,1]}\theta_k\left( \frac{x + (s - 1)y}{s}, t\right)\frac{\Delta x}{s}P_k(y,t) \Delta x \, \mathrm{d} y   \,.
\end{align}


\subsubsection{Expected number of degree-$k$ nodes with opinions that leave the interval $[x, x + \Delta x)$ at time $t + 1$}
We now consider the mean number of degree-$k$ nodes whose opinion is in $[x, x + \Delta x)$ at time $t$ but not at time $t + 1$.
We argue as in \cref{App:sec:mean_enter_interval}, although now our argument is a bit simpler.
Because the opinion of a degree-$k$ node follows the distribution $P_k(x,t)$ and the weighted median of its neighbors' opinions has the distribution $\theta_k(x,t)$ in the limit $N \rightarrow \infty$, the probability that $x_i(t)\in[x, x + \Delta x)$ and $x_i(t + 1)\notin [x, x + \Delta x)$ in this limit is
\begin{align}
    \left(1- \theta_k(x,t) \Delta x\right)P_k(x,t) \Delta x \,.
\end{align}
The mean number of such nodes is thus
\begin{align}\label{Theory:eq:mean_field_7}
    N q_k \left(1 - \theta_k(x,t) \Delta x\right)P_k(x,t) \Delta x \,.
\end{align}


\subsubsection{Expected change in the number of degree-$k$ nodes with opinions in the interval $[x, x + \Delta x)$ at time $t + 1$}
Combining equations \cref{Theory:eq:mean_field_1}, \cref{Theory:eq:mean_field_8}, and \cref{Theory:eq:mean_field_7} yields
\begin{align}
    &Nq_k P_k(x,t + 1) \Delta x - Nq_k P_k(x,t) \Delta x 
     \\
    &\qquad = N q_k \int_{y\in [0,1]}\theta_k\left( \frac{x + (s-1) y}{s}, t\right)\frac{\Delta x}{s}P_k(y,t) \, \mathrm{d} y \Delta x - N q_k \left(1 - \theta_k(x,t)\Delta x\right)P_k(x,t) \Delta x\,, \nonumber 
\end{align}
which implies that
\begin{align}\label{Theory:eq:mean_field_9}
    P_k(x,t + 1) 
    &\rightarrow 
    \frac{1}{s}\int_{y\in [0,1]}\theta_k\left( \frac{x + (s - 1)y}{s}, t\right)P_k(y,t) \, \mathrm{d} y \quad \text{as} \quad \Delta x \rightarrow 0\, .
\end{align}
This completes our derivation of the mean-field approximation {(\ref{Theory:eq:mean_field_eq}, \ref{Theory:eq:global_dens})}.


\subsection{Finite-size effects and inaccuracies from annealing-assumption effects}\label{App:finite-size}
In our derivation of the mean-field approximation {(\ref{Theory:eq:mean_field_eq}, \ref{Theory:eq:global_dens})}, we made two key assumptions: (1) we supposed that the network size (i.e., number of nodes) $N \rightarrow \infty$; 
 and (2) we assumed that the dynamics occur on an ``annealed" network \cite{Dorogovtsev2008CriticalNetworks}. 
Both of these choices affect the accuracy of our approximation.

We now examine finite-size effects (i.e., inaccuracies that arise from our $N \rightarrow \infty$ asymptotic assumption) and annealing-assumption effects (i.e., inaccuracies that arise from the annealed-network assumption) in our mean-field approximation. Both of these effects occur, but they seem to diminish quickly with time for most values of self-appraisal.
We thus conclude that both effects are {eventually} dominated by the impact of self-appraisal on the accuracy of our mean-field approximation.

To examine finite-size effects, we calculate the root-mean-square error (RMSE) between numerical solutions of our mean-field approximation {(\ref{Theory:eq:mean_field_eq}, \ref{Theory:eq:global_dens})} and the means {of $100$} simulations of our weighted-median opinion model {(\ref{eq:WMMI}, \ref{Theory:def:weighted_median})} for different network sizes $N$.
We choose the initial opinions uniformly at random.

To examine annealing-assumption effects, we calculate means of simulations of our weighted-median {opinion} model {(\ref{eq:WMMI}, \ref{Theory:def:weighted_median})} on both time-independent (i.e., ``fixed") networks and time-dependent (i.e., ``annealed'') networks.
For both types of networks, we start with a $(k_1, k_2)$-regular configuration-model network with degrees $k_1 = 11$ and $k_2 = 101$ and associated probabilities $q_{k_1} = 0.9$ and $q_{k_2} = 0.1$. 
A {fixed} network is a network that never changes with time.
By contrast, an {annealed} network changes with time.
At each discrete time step, we uniformly randomly rewire the edges while preserving the network's degree sequence. 
That is, for each time step, the annealed network is one instantiation of the selected $(k_1, k_2)$-regular configuration-model ensemble.
Therefore, the annealed networks satisfy the annealing assumption in our mean-field approximation (\ref{Theory:eq:mean_field_eq}, \ref{Theory:eq:global_dens}), whereas the fixed networks do not satisfy this assumption.

In \cref{Res:fig:MF_finite_size_config}, we show the RMSEs from these numerical experiments and explore how they change as time increases from $t = 1$ to $t = 5$.
At time $t = 1$ and for all values of the self-appraisal $s$, the RMSE is smaller for larger networks for both the annealed and the fixed networks.
Notably, for $t = 1$, network size seems to affect the RMSE more than the self-appraisal value. 
The RMSE has a similar magnitude for both the annealed and fixed networks.
Therefore, for $t = 1$, it seems that finite-size effects impact the accuracy of our mean-field approximation 
but that there are not any significant annealing-assumption effects.

At time $t = 3$, finite-size effects still impact the accuracy of the mean-field approximation{, but these effects are smaller than the} effects of the self-appraisal value.
Additionally, the annealing-assumption effects are more pronounced for $t = 3$ than for $t = 1$.
At time $t = 5$, the finite-size effects are still noticeable for the annealed networks for $s \leq 0.5$, but they are barely noticeable for the fixed networks.
{For $t = 3$ and $t = 5$,} the self-appraisal has a larger impact than either finite-size effects or annealing-assumption effects on the accuracy of our mean-field approximation.

\begin{figure}[h!]
    \centering
   \includegraphics[width  = 0.95 \textwidth]{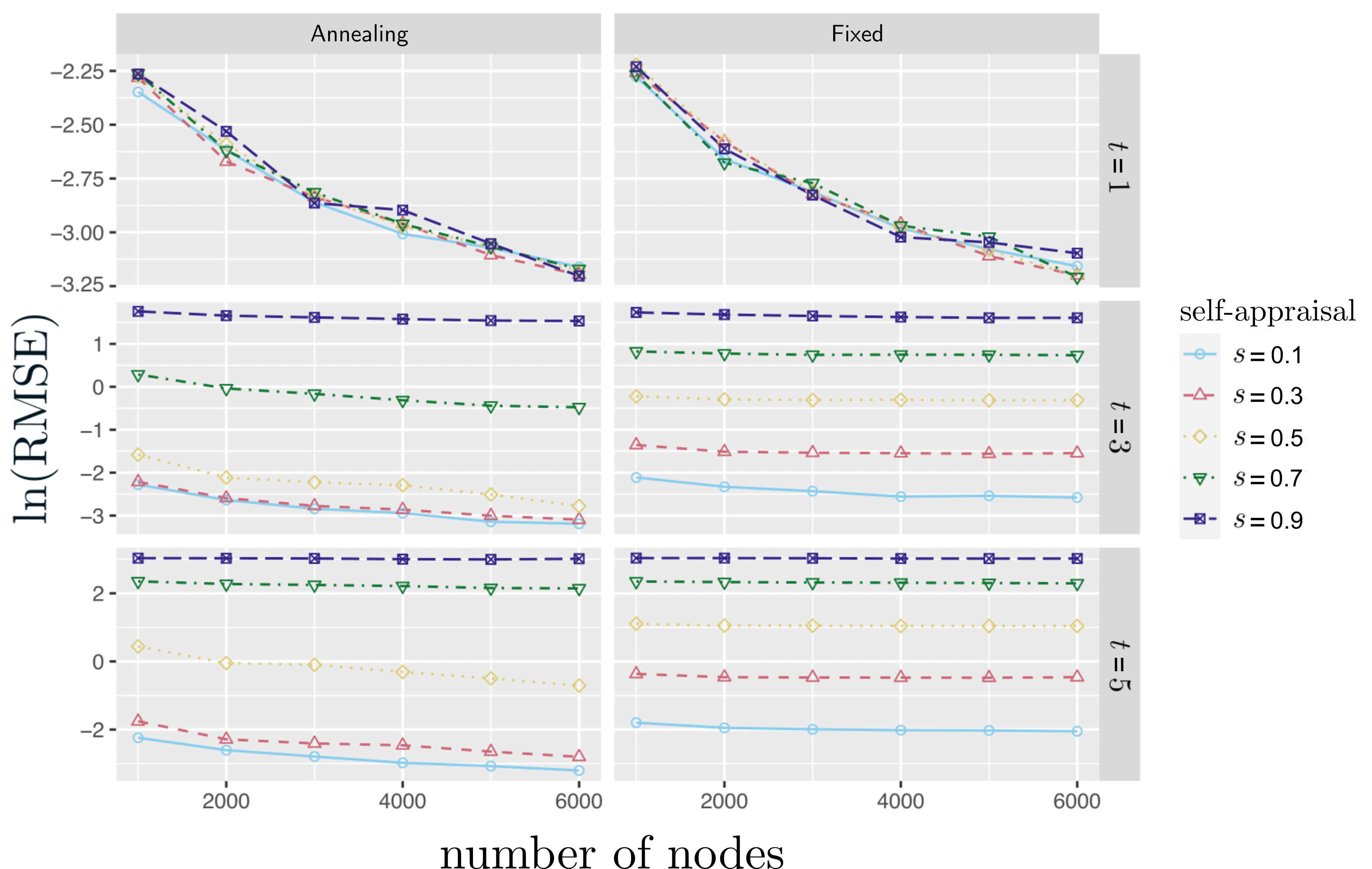}
    \caption{
        The natural logarithm of the RMSE between our weighted-median opinion model (\ref{eq:WMMI}, \ref{Theory:def:weighted_median}) and its mean-field approximation (\ref{Theory:eq:mean_field_eq}, \ref{Theory:eq:global_dens}) for annealed and {fixed} networks that we generate using a $(k_1, k_2)$-regular configuration model with different numbers of nodes.
        The degrees are $k_1 = 11$ and $k_2 = 101$, and the associated probabilities are $q_{k_1} = 0.9$ and $q_{k_2} = 0.1$.
        At time $t = 1$, the RMSE is smaller for larger networks for all values of the self-appraisal $s$.
        However, at later times, the network size has less influence.
        At time $t = 5$, the effect of self-appraisal dominates both annealing-assumption effects and finite-size effects.
    }
    \label{Res:fig:MF_finite_size_config}
\end{figure}


\subsection{Effect of self-appraisal on the root-mean-square error (RMSE) between our weighted-median {opinion} model (\ref{eq:WMMI}, \ref{Theory:def:weighted_median}) and its mean-field approximation (\ref{Theory:eq:mean_field_eq}, \ref{Theory:eq:global_dens})}\label{App:sec:MF_RMSE_refined}
In \cref{Res:fig:MF_accuracy_Bowdoin_fine_grined}, we show the dependence of the RMSE between our weighted-median {opinion} model (\ref{eq:WMMI}, \ref{Theory:def:weighted_median}) and its mean-field approximation {(\ref{Theory:eq:mean_field_eq}, \ref{Theory:eq:global_dens})} for the Bowdoin Facebook friendship network using a more finely-gained set of self-appraisal values than the ones that we used in section \ref{sec:num_test_MF}.
For self-appraisal $s = 0.1$, the RMSE initially increases rapidly and it then increases slowly for $t \geq 1$.
For $s = 0.3$, the RMSE initially increases slowly and it then increases much more rapidly.
By examining self-appraisal values between these extremes, we observe a seemingly smooth transition between these two behaviors.
We observe similar qualitative features for our other real-world social networks.

\begin{figure}[h!]
    \centering
    \includegraphics[width = 0.95 \textwidth]{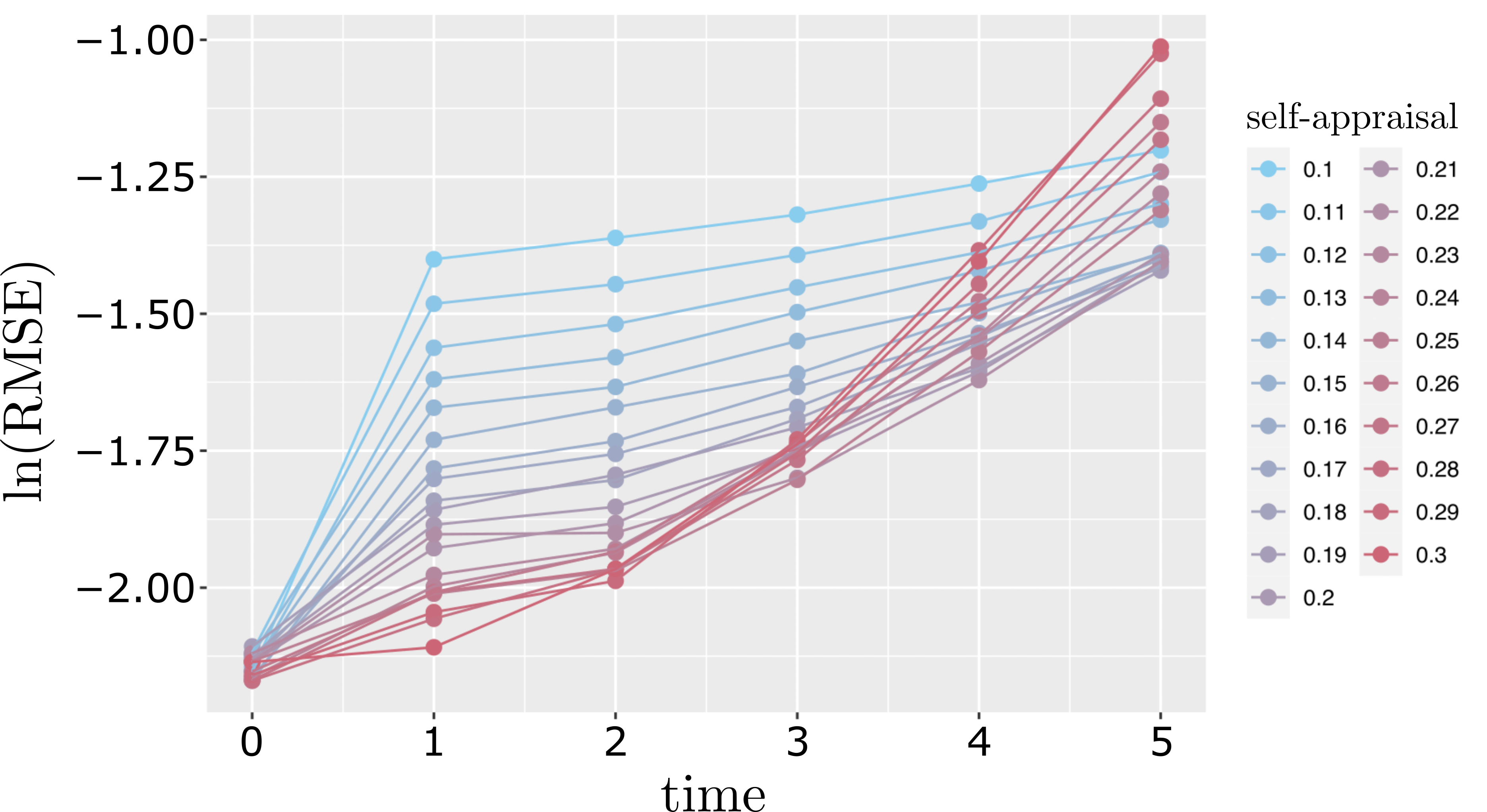}
    \caption{
    The natural logarithm of the RMSE between our weighted-median opinion model {(\ref{eq:WMMI}, \ref{Theory:def:weighted_median})} and its mean-field approximation {(\ref{Theory:eq:mean_field_eq}, \ref{Theory:eq:global_dens})} for the Bowdoin Facebook friendship network for self-appraisal values between 
    $s = 0.1$ and $s = 0.3$. The RMSE increases rapidly at earlier times for self-appraisal values near $0.1$ than it does for self-appraisal values near $0.3$.
    There appears to be a smooth transition between these two behaviors.
    }
    \label{Res:fig:MF_accuracy_Bowdoin_fine_grined}
\end{figure}


\subsection{Effect of degree ``lumping'' on the RMSE between our weighted-median {opinion} model (\ref{eq:WMMI}, \ref{Theory:def:weighted_median}) and its mean-field approximation (\ref{Theory:eq:mean_field_eq}, \ref{Theory:eq:global_dens})}\label{sec:lumping}
We observed in \cref{sec:num_test_MF} that the maximum value of $\theta_k$ in our mean-field approximation {(\ref{Theory:eq:mean_field_eq}, \ref{Theory:eq:global_dens})} increases with the node degree $k$.
This results in numerical overflow in the evaluation of $\theta_k$ for sufficiently large values of $k$.
To deal with this issue, we introduce a ``lumping degree'' and aggregate all degree-$k$ nodes when $k$ is at least that degree into a single degree class with a common opinion distribution.
The probability that a node is in the lumped degree class is the sum of the probabilities that it is any of its constituent degree classes.

Using numerical computations, we examine the effect of different lumping degrees on the mean-field approximation {(\ref{Theory:eq:mean_field_eq}, \ref{Theory:eq:global_dens})}.
In \cref{Res:fig:MF_lumping_Georgetown}, we show the RMSE between our weighted-median opinion model {(\ref{eq:WMMI}, \ref{Theory:def:weighted_median})} and its mean-field approximation {(\ref{Theory:eq:mean_field_eq}, \ref{Theory:eq:global_dens})} for the Georgetown Facebook network at times $t = 1$ and $t = 4$ for different lumping degrees and self-appraisal values. 
The behavior of the RMSE depends on the self-appraisal $s$.
At time $t = 4$, for self-appraisal $s = 0.1$, the RMSE initially decreases with the lumping degree.
However, the RMSE initially increases with lumping degree for larger self-appraisal values.

In \cref{Res:fig:MF_lumping_Georgetown_fine_grained_full}, we show the RMSE between our weighted-median {opinion} model its the mean-field approximation {(\ref{Theory:eq:mean_field_eq}, \ref{Theory:eq:global_dens})} for the Georgetown Facebook network at times $t = 1$ and $t = 4$ for different lumping degrees for self-appraisals $s \in [0.1,0.3]$.
There appears to be a smooth transition between the observed behaviors.

Based on our simulations of our mean-field approximation {(\ref{Theory:eq:mean_field_eq}, \ref{Theory:eq:global_dens})} on the Facebook friendship and Twitter followership networks, there also appears to be little effect of increasing the lumping degree beyond $500$. For all of the results in \cref{sec:MF_acc}, we use a lumping degree of $1000$.

\begin{figure}[h!]
    \centering
    \includegraphics[width  = 0.9 \textwidth]{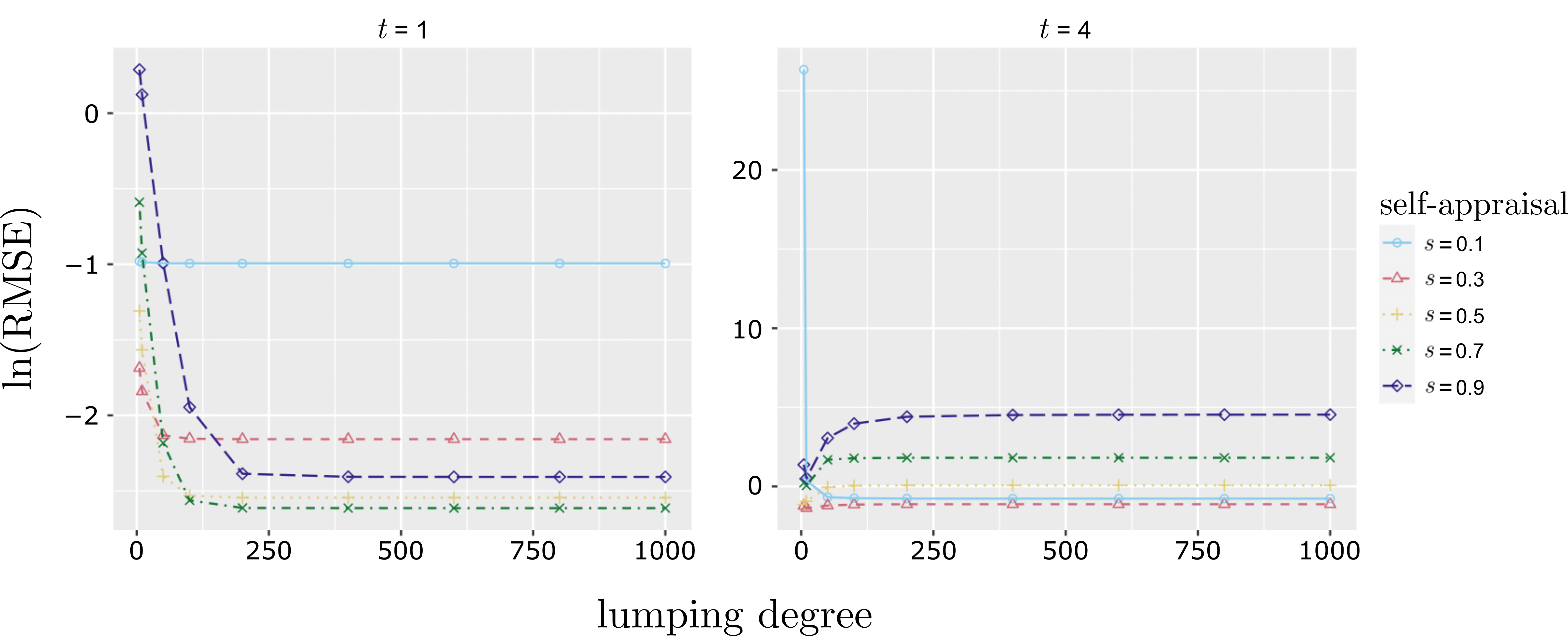}
    \caption{
    he natural logarithm of the RMSE between our weighted-median opinion model {(\ref{eq:WMMI}, \ref{Theory:def:weighted_median})} and its mean-field approximation {(\ref{Theory:eq:mean_field_eq}, \ref{Theory:eq:global_dens})} for the Georgetown Facebook friendship network at times $t = 1$ and $t = 4$ for various lumping degrees and self-appraisal values $s$.
    The RMSE is similar for all lumping degrees of at least $500$. 
    This is the case for all of the Facebook friendship networks and Twitter followership networks.
    }
    \label{Res:fig:MF_lumping_Georgetown}
\end{figure}

\begin{figure}[h!]
    \centering
    \includegraphics[width  = 0.95 \textwidth]{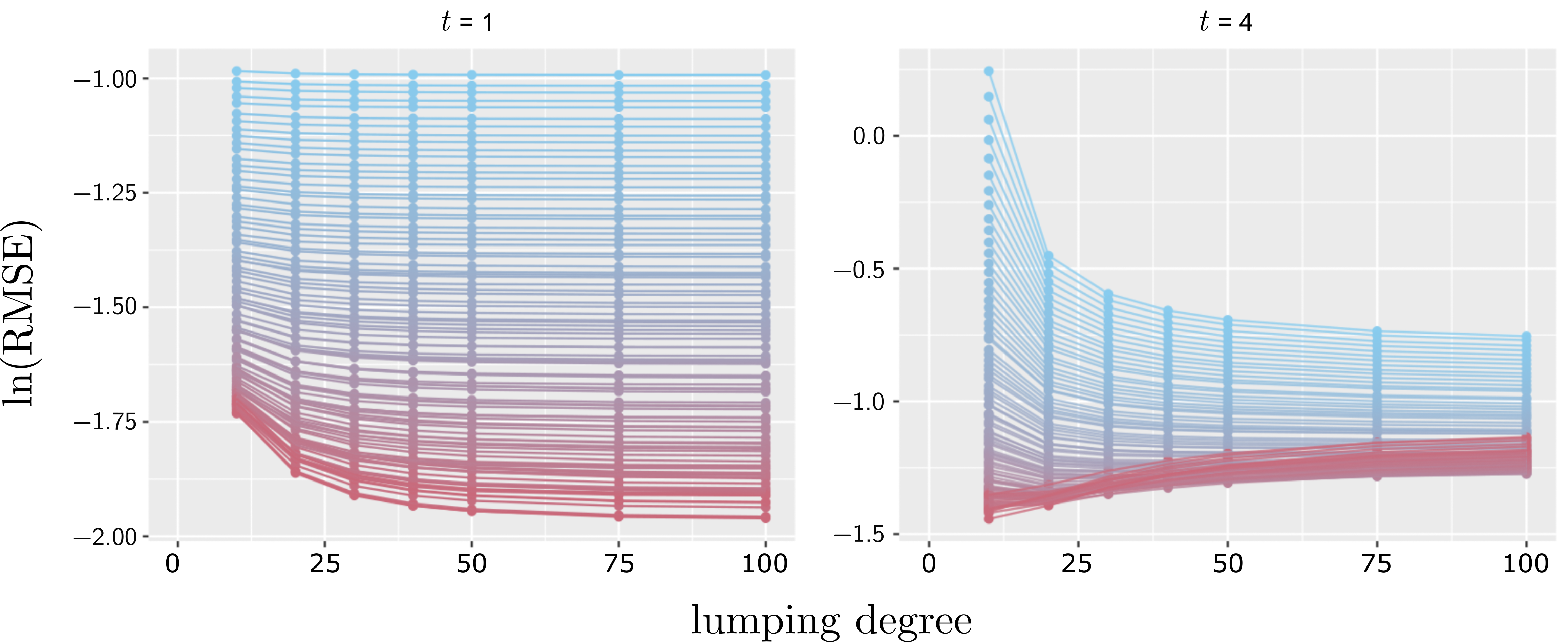}
    \caption{
    The natural logarithm of the RMSE between our weighted-median opinion model (\ref{eq:WMMI}, \ref{Theory:def:weighted_median}) and its mean-field approximation (\ref{Theory:eq:mean_field_eq}, \ref{Theory:eq:global_dens}) for the Georgetown Facebook friendship network at times $t = 1$ and $t = 4$ for various lumping degrees and self-appraisal values ranging from $s = 0.1$ (light blue) to $s = 0.3$ (red).
    The RMSE evolves differently for $s = 0.1$ than it does for $s = 0.3$.
    At $t = 1$, the RMSE depends little on the lumping degree for $s = 0.1$, but it decreases for larger lumping degrees for $s = 0.3$. At $t = 4$, the RMSE decreases for larger lumping degrees for $s = 0.1$, but it increases for larger lumping degrees for $s = 0.3$. There appears to be a smooth transition between these the two regimes.
    }\label{Res:fig:MF_lumping_Georgetown_fine_grained_full}
\end{figure}


\section*{Acknowledgements}

We thank Christian Henriksen for helpful discussions.


\bibliographystyle{siamplain}
\bibliography{references-v15}

\begin{thebibliography}{10}

\bibitem{granov2017VotersTurnOut}
{\sc M.~Agranov, J.~K. Goeree, J.~Romero, and L.~Yariv}, {\em What makes voters turn out: {T}he effects of polls and beliefs}, Journal of the European Economic Association, 16 (2017), pp.~825--856.

\bibitem{Barabasi1999}
{\sc A.~L. Barab{\'{a}}si and R.~Albert}, {\em {Emergence of scaling in random networks}}, Science, 286 (1999), pp.~509--512.

\bibitem{Baron2021ConsensusDisorder}
{\sc J.~W. Baron}, {\em Consensus, polarization, and coexistence in a continuous opinion dynamics model with quenched disorder}, Physical Review E, 104 (2021), 044309.

\bibitem{Beasley2001}
{\sc R.~K. Beasley and M.~R. Joslyn}, {\em Cognitive dissonance and post-decision attitude change in six presidential elections}, Political Psychology, 22 (2001), pp.~521--540.

\bibitem{Bedson2021SocialDiseaseModel}
{\sc J.~Bedson, L.~A. Skrip, D.~Pedi, S.~Abramowitz, S.~Carter, M.~F. Jalloh, S.~Funk, N.~Gobat, T.~Giles-Vernick, G.~Chowell, J.~R. de~Almeida, R.~Elessawi, S.~V. Scarpino, R.~A. Hammond, S.~Briand, J.~M. Epstein, L.~H{\'e}bert-Dufresne, and B.~M. Althouse}, {\em A review and agenda for integrated disease models including social and behavioural factors}, Nature Human Behaviour, 5 (2021), pp.~834--846.

\bibitem{bernardo2024}
{\sc C.~Bernardo, C.~Altafini, A.~Proskurnikov, and F.~Vasca}, {\em Bounded confidence opinion dynamics: {A} survey}, Automatica, 159 (2024), 111302.

\bibitem{Biddlestone2020ConspiracyCovid}
{\sc G.~Biddlestone and K.~M. Douglas}, {\em Cultural orientation, power, belief in conspiracy theories, and intentions to reduce the spread of {COVID-19}}, British Journal of Social Psychology, 59 (2020), pp.~663--673.

\bibitem{BrooksIdeologyContent2020}
{\sc H.~Z. Brooks and M.~A. Porter}, {\em A model for the influence of media on the ideology of content in online social networks}, Physical Review Research, 2 (2020), 023041.

\bibitem{Caron-Lormier2008}
{\sc G.~Caron-Lormier, R.~W. Humphry, D.~A. Bohan, C.~Hawes, and P.~Thorbek}, {\em {Asynchronous and synchronous updating in individual-based models}}, Ecological Modelling, 212 (2008), pp.~522--527.

\bibitem{Sarndal2004OrderStatistics}
{\sc H.~A. David and H.~N. Nagaraja}, {\em Order Statistics}, John Wiley {\&} Sons, Inc., New York City, NY, USA, 3rd~ed., 2003.

\bibitem{Degroot1974}
{\sc M.~H. DeGroot}, {\em Reaching a consensus}, Journal of the American Statistical Association, 69 (1974), pp.~118--121.

\bibitem{Dorogovtsev2008CriticalNetworks}
{\sc S.~N. Dorogovtsev, A.~V. Goltsev, and J.~F.~F. Mendes}, {\em Critical phenomena in complex networks}, Reviews of Modern Physics, 80 (2008), pp.~1275--1335.

\bibitem{EEti2020Centrality-based}
{\sc Eeti, A.~Singh, and H.~Cherifi}, {\em Centrality-based opinion modeling on temporal networks}, IEEE Access, 8 (2020), pp.~1945--1961.

\bibitem{fennell2021}
{\sc S.~C. Fennell, K.~Burke, M.~Quayle, and J.~P. Gleeson}, {\em Generalized mean-field approximation for the {D}effuant opinion dynamics model on networks}, Physical Review E, 103 (2021), 012314.

\bibitem{Festinger1957}
{\sc L.~Festinger}, {\em {A Theory of Cognitive Dissonance}}, Stanford University Press, Stanford, CA, USA, 1957.

\bibitem{Festinger1959a}
{\sc L.~Festinger and J.~M. Carlsmith}, {\em {Cognitive consequences of forced compliance}}, Journal of Abnormal and Social Psychology, 58 (1959), pp.~203--210.

\bibitem{Fosdick2016ConfiguringSequences}
{\sc B.~Fosdick, D.~Larremore, J.~Nishimura, and J.~Ugander}, {\em Configuring random graph models with fixed degree sequences}, SIAM Review, 60 (2018), pp.~315--355.

\bibitem{friedkin2014}
{\sc N.~E. Friedkin and E.~C. Johnsen}, {\em Social Influence Network Theory: A Sociological Examination of Small Group Dynamics}, Cambridge University Press, Cambridge, UK, 2014.

\bibitem{galesic2021}
{\sc M.~Galesic, H.~Olsson, J.~Dalege, T.~van~der Does, and D.~L. Stein}, {\em Integrating social and cognitive aspects of belief dynamics: {T}owards a unifying framework}, J. R. Soc. Interface, 18 (2021), 20200857.

\bibitem{Garimella2018}
{\sc K.~Garimella, G.~De~Francisci~Morales, A.~Gionis, and M.~Mathioudakis}, {\em Political discourse on social media: Echo chambers, gatekeepers, and the price of bipartisanship}, in Proceedings of the 2018 World Wide Web Conference, WWW '18, Republic and Canton of Geneva, Switzerland, 2018, pp.~913--922.

\bibitem{Han2024continuoustime}
{\sc Y.~Han, G.~Chen, F.~D\"{o}rfler, and W.~Mei}, {\em The continuous-time weighted-median opinion dynamics}, 2024, \url{https://arxiv.org/abs/2404.16318}.

\bibitem{HickockOpinDynHypergraphs2022}
{\sc A.~Hickok, Y.~Kureh, H.~Z. Brooks, M.~Feng, and M.~A. Porter}, {\em A bounded-confidence model of opinion dynamics on hypergraphs}, SIAM Journal on Applied Dynamical Systems, 21 (2022), pp.~1--32.

\bibitem{iacopini2019}
{\sc I.~Iacopini, G.~Petri, A.~Barrat, and V.~Latora}, {\em Simplicial models of social contagion}, Nature Communications, 10 (2019), 2485.

\bibitem{Jia2015}
{\sc P.~Jia, A.~Mirtabatabaei, N.~E. Friedkin, and F.~Bullo}, {\em Opinion dynamics and the evolution of social power in influence networks}, SIAM Review, 57 (2015), pp.~367--397.

\bibitem{Lazarsfeld}
{\sc P.~F. Lazarsfeld, B.~Berelson, and H.~Gaudet}, {\em {The People's Choice. How the Voter Makes Up His Mind in a Presidential Campaign: Legacy edition}}, Columbia University Press, New York City, NY, USA, 2021.

\bibitem{lee2024}
{\sc M.~Lee}, {\em Is polarization an inevitable outcome of similarity-based content recommendations? --- {M}athematical proofs and computational validation}, 2024, \url{https://arxiv.org/abs/2412.10524}.

\bibitem{Leon-Medina2020Fakers}
{\sc F.~J. Le\'{o}n-Medina, J.~Tena-S\'{a}nchez, and F.~J. Miguel}, {\em Fakers becoming believers: How opinion dynamics are shaped by preference falsification, impression management and coherence heuristics}, Quality \& Quantity, 54 (2020), p.~385–412.

\bibitem{LiWeightedMedian2022}
{\sc G.~Li, Q.~Liu, and L.~Chai}, {\em Analysis and application of weighted-median {H}egselmann--{K}rause opinion dynamics model on social networks}, in 2022 34th Chinese Control and Decision Conference (CCDC), 2022, pp.~5409--5414.

\bibitem{Lorenz2006ConsensusConfidence}
{\sc J.~Lorenz}, {\em Consensus strikes back in the {Hegselmann--Krause} model of continuous opinion dynamics under bounded confidence}, Journal of Artificial Societies and Social Simulation, 9 (2006), 8.

\bibitem{Lorenz2010HeterogeneousConsensus}
{\sc J.~Lorenz}, {\em Heterogeneous bounds of confidence: {M}eet, discuss and find consensus!}, Complexity, 15 (2010), pp.~43--52.

\bibitem{Lorenz2010Convergence}
{\sc J.~Lorenz and D.~A. Lorenz}, {\em On conditions for convergence to consensus}, IEEE Transactions on Automatic Control, 55 (2010), pp.~1651--1656.

\bibitem{mas2019}
{\sc M.~M{\"a}s}, {\em Challenges to simulation validation in the social sciences. {A} critical rationalist perspective}, in Computer Simulation Validation: Fundamental Concepts, Methodological Frameworks, and Philosophical Perspectives, C.~Beisbart and N.~J. Saam, eds., Springer International Publishing, Cham, Switzerland, 2019, pp.~857--879.

\bibitem{Matz2005a}
{\sc D.~C. Matz and W.~Wood}, {\em {Cognitive dissonance in groups: The consequences of disagreement}}, Journal of Personality and Social Psychology, 88 (2005), pp.~22--37.

\bibitem{MedoFragility2021}
{\sc M.~Medo, M.~S. Mariani, and L.~L\"{u}}, {\em The fragility of opinion formation in a complex world}, Communications Physics, 4 (2021), 75.

\bibitem{MeiMicro-Foundation2022}
{\sc W.~Mei, F.~Bullo, G.~Chen, J.~M. Hendrickx, and F.~D\"orfler}, {\em Micro-foundation of opinion dynamics: Rich consequences of the weighted-median mechanism}, Physical Review Research, 4 (2022), 023213.

\bibitem{mei2022convergence}
{\sc W.~Mei, J.~M. Hendrickx, G.~Chen, F.~Bullo, and F.~D\"{o}rfler}, {\em Convergence, consensus and dissensus in the weighted-median opinion dynamics}, IEEE Transactions on Automatic Control, 69 (2024), pp.~6700--6714.

\bibitem{melnik2013}
{\sc S.~Melnik, J.~A. Ward, J.~P. Gleeson, and M.~A. Porter}, {\em Multi-stage complex contagions}, Chaos, 23 (2013), 013124.

\bibitem{Meng2018OpinionNetworks}
{\sc X.~F. Meng, R.~A.~V. Gorder, and M.~A. Porter}, {\em {Opinion formation and distribution in a bounded-confidence model on various networks}}, Physical Review E, 97 (2018), 022312.

\bibitem{newman2018}
{\sc M.~E.~J. Newman}, {\em Networks}, Oxford University Press, Oxford, UK, second~ed., 2018.

\bibitem{Noorazar2020FromDynamics}
{\sc H.~Noorazar, K.~R. Vixie, A.~Talebanpour, and Y.~Hu}, {\em From classical to modern opinion dynamics}, International Journal of Modern Physics C, 31 (2020), 2050101.

\bibitem{Pitman1993Probability}
{\sc J.~Pitman}, {\em {Probability}}, Springer-Verlag, Heidelberg, Germany, 1993.

\bibitem{PorterGleeson2016}
{\sc M.~A. Porter and J.~P. Gleeson}, {\em Dynamical Systems on Networks: A Tutorial}, vol.~4 of Frontiers in Applied Dynamical Systems: Reviews and Tutorials, Springer International Publishing, Cham, Switzerland, 2016.

\bibitem{Redner2019Reality-inspired}
{\sc S.~Redner}, {\em Reality-inspired voter models: {A} mini-review}, Comptes Rendus Physique, 20 (2019), pp.~275--292.

\bibitem{Schawe2021WhenSocieties}
{\sc H.~Schawe, S.~Fontaine, and L.~Hern{\'{a}}ndez}, {\em {When network bridges foster consensus. {B}ounded confidence models in networked societies}}, Physical Review Research, 3 (2021), 023208.

\bibitem{starnini2025}
{\sc M.~Starnini, F.~Baumann, T.~Galla, D.~Garcia, G.~I{\~n}iguez, M.~Karsai, J.~Lorenz, and K.~Sznajd-Weron}, {\em Opinion dynamics: {S}tatistical physics and beyond}, arXiv:2507.11521,  (2024).

\bibitem{KashinMasonTieDecay}
{\sc K.~Sugishita, M.~A. Porter, M.~Beguerisse-D\'{\i}az, and N.~Masuda}, {\em Opinion dynamics on tie-decay networks}, Physical Review Research, 3 (2021), 023249.

\bibitem{TormalaAttituide2018}
{\sc Z.~L. Tormala and D.~D. Rucker}, {\em Attitude certainty: {A}ntecedents, consequences, and new directions}, Consumer Psychology Review, 1 (2018), pp.~72--89.

\bibitem{Traud2012}
{\sc A.~L. Traud, P.~J. Mucha, and M.~A. Porter}, {\em {Social structure of {F}acebook networks}}, Physica A: Statistical Mechanics and its Applications, 391 (2012), pp.~4165--4180.

\bibitem{Kerckhove2016}
{\sc C.~Vande~Kerckhove, S.~Martin, P.~Gend, P.~J. Rentfrow, J.~M. Hendrickx, and V.~D. Blondel}, {\em Modelling influence and opinion evolution in online collective behaviour}, PLOS ONE, 11 (2016), pp.~1--25.

\bibitem{Watts1998}
{\sc D.~J. Watts and S.~H. Strogatz}, {\em {Collective dynamics of `small-world' networks}}, Nature, 393 (1998), pp.~440--442.

\bibitem{Xia2011OpinionDynamics}
{\sc H.~Xia, H.~Wang, and Z.~Xuan}, {\em Opinion dynamics: {A} multidisciplinary review and perspective on future research}, International Journal of Knowledge and Systems Science, 2 (2011), pp.~72--91.

\bibitem{ZhangWMMPrejudice2025}
{\sc R.~Zhang, Z.~Liu, G.~Chen, and W.~Mei}, {\em Convergence analysis of weighted-median opinion dynamics with prejudice}, IEEE Transactions on Automatic Control, 70 (2025), pp.~4155--4162.

\bibitem{DingReachingConsensus2019}
{\sc D.~Zhaogang, C.~Xia, D.~Yucheng, and F.~Herrera}, {\em Consensus reaching in social network {DeGroot} model: {T}he roles of the self-confidence and node degree}, Information Sciences, 486 (2019), pp.~62--72.

\end{thebibliography}


\end{document}